\documentclass[Crown,sagev,times]{sagej}
\usepackage{moreverb,url}

\usepackage[colorlinks,bookmarksopen,bookmarksnumbered,citecolor=red,urlcolor=red]{hyperref}

\usepackage{color}
\usepackage{lscape,rotating}
\usepackage{float,epsfig}
\usepackage{subfigure}
\usepackage{graphicx}
\usepackage{setspace}
 \usepackage{amsmath,bm} 
\usepackage{amsfonts}
\usepackage{amssymb}

\newtheorem{lemma}{Lemma}

\newcommand{\muv}{\bm{\mu}}
\newcommand{\zerov}{\bm{0}}

\newcommand{\xv}{\bm{x}}
\newcommand{\yv}{\bm{y}}
\newcommand{\zv}{\bm{z}}
\newcommand{\Ev}{\bm{E}}

\newcommand{\wc}{\text{\small$\mathcal{W}$}}
\newcommand{\ps}{\text{\tiny$\psi$}}

\newcommand{\thetav}{\bm{\theta}}

\newcommand{\phiv}{\bm{\phi}}

\newcommand{\xiv}{\bm{\xi}}
\newcommand{\psiv}{\bm{\psi}}
\newcommand{\alphav}{\bm{\alpha}}

\newcommand{\lambdav}{\bm{\lambda}}
\newcommand{\etav}{\bm{\eta}}

\newcommand{\epsilonv}{\bm{\epsilon}}
\newcommand{\Qss}{\text{\small{$Q$}}}
\newcommand{\Qs}{\text{\tiny{$Q$}}}
\newcommand{\Rss}{\text{\small{$R$}}}
\newcommand{\Rs}{\text{\tiny{$R$}}}

\begin{document}

\runninghead{MDA algorithm for multivariate skew-t regression}

\title{A monotone data augmentation algorithm for longitudinal data analysis via multivariate skew-t, skew-normal or t distributions}

\author{Yongqiang Tang}

\affiliation{Tesaro, 1000 Winter Street, Waltham, MA 02451, USA}

\corrauth{Yongqiang Tang}

\email{yongqiang\_tang@yahoo.com}

\begin{abstract}
The mixed effects model for repeated measures (MMRM) has been widely used for the analysis of longitudinal clinical data collected at a number of fixed time points.
We propose a robust extension of the MMRM  for skewed and heavy-tailed data on basis of the multivariate skew-t distribution,
and it includes the multivariate normal, t, and skew-normal distributions as special cases. An efficient Markov chain Monte Carlo  algorithm is developed using the
monotone data augmentation and parameter expansion techniques.
We employ the algorithm to perform controlled pattern imputations for sensitivity analyses of longitudinal clinical trials with nonignorable dropouts. The proposed methods
are illustrated by real data analyses. Sample SAS programs for the analyses are provided in the online supplementary material.
\end{abstract}


\maketitle

\section{Introduction}\label{intsec}

The mixed effects model for repeated measures (MMRM) has been
commonly used for the primary analysis of longitudinal continuous outcomes in clinical trials \cite{aiddiqui:2009,mallinckrodt:2008}. 
As recommended by the regulatory guidelines  \cite{e9:1999,chmp:2015}, the main analysis in clinical trials shall be unambiguously prespecified  in the protocol.
For this reason,  the observations within a subject are usually 
assumed to follow a multivariate normal distribution with an unstructured covariance matrix  in MMRM \cite{aiddiqui:2009,mallinckrodt:2008, laird:1987, tang:2017sim}.
If  the covariance  structure is misspecified, the treatment effect estimate may be biased in the presence of missing data \cite{lu:2010},
and the test for the fixed effects may not be able to control the type I error rate  \cite{gurka:2011,gadda:2007}. A covariance selection approach using Akaike’s information criterion or Schwarz’s Bayesian
information criterion also tends to inflate the type I error rate \cite{gomez:2005}.

The controlled imputation methodology \cite{little:1996, 2013:carpenter, 2013:mallinckrodta, 2017:tangb} has become increasingly popular in
 sensitivity analyses of longitudinal trials with nonignorable dropouts. The controlled imputation and MMRM assume the same observed  data distribution, but specify  different  mechanisms for missingness due to dropout.
The  MMRM assumes  the  data  are missing at random (MAR \cite{rubin:1976}). It implies that subjects who discontinue  the treatment early  have the same mean response profiles as subjects who complete the  trial. 
The MAR mechanism may not be convincing if the dropout is due to adverse events or inadequate efficacy.
Sensitivity analyses under missing not at random (MNAR \cite{rubin:1976}) are recommended by recent regulatory guidelines \cite{ich:2017,chmp:2010} and a FDA-mandated panel report from the National Research Council \cite{NRC:2010}.
Under MNAR, the  response profiles differ systematically between  dropouts and completers. 
In the controlled imputation, the mean outcomes among subjects in the experimental arm after dropout are assumed to be similar to that of control subjects, or get worse compared to subjects  who stay on the  treatment.
The assumption is clinically plausible and easy to understand.
If the treatment effect remains significant in a conservative nonignorable sensitivity analysis, we can claim the robustness of the primary conclusion obtained under MAR.

The controlled imputation  specifies a pattern mixture model (PMM \cite{little:1993}) for the longitudinal outcomes in the sense that  their joint distribution depends on the dropout time. 
The controlled imputation is generally implemented via multiple imputation (MI), and this will be explained in Section \ref{discussion}. Tang \cite{2017:tangb,2017:tanga} introduces a formal Markov chain Monte Carlo (MCMC) algorithm to conduct the controlled imputation based on the   monotone data augmentation (MDA) strategy.
It  extends and improves Schafer's \cite{schafer:1997} MDA algorithm for multivariate normal outcomes. 
The algorithm is unaffected by the dropout  mechanism. The missing data after dropout  are integrated out  of the posterior distribution, and imputed after  the algorithm converges.
Only the intermittent missing outcomes and the model parameters are drawn
iteratively  before  the Markov chain reaches its stationary distribution.  It imputes fewer missing values, and tends to converge faster with smaller autocorrelation between
posterior samples than an algorithm that imputes all missing outcomes in each  iteration \cite{schafer:1997, 2017:tanga}. 
Tang \cite{2018:tangd, tang:2019} develops the controlled imputation  for longitudinal outcomes with potentially different types of variables based on the factored likelihood,
in which the conditional model of the outcome at each visit given the outcomes at previous visits can be  
linear, binary logistic, multinomial logistic, proportional odds, Poisson, negative binomial, skew-normal (SN \cite{AZZALINI:1985}) or skew-t (ST \cite{azzalini:2003}) regressions, and may vary by visits.

The main purpose of this article is to extend the controlled imputation  to non-normal longitudinal continuous outcomes. In the mixed effects model,
inference about the fixed effects is asymptotically valid for non-normal outcomes  \cite{verbeke:1997}, but may suffer from some loss of efficiency \cite{pinheiro:2001, zhang:2001} in large samples.
 The inference is  vulnerable to severe departures from normality  in small and moderate samples \cite{arnau:2013}, and  
this can be easily understood in the simple case of a  t test \cite{boos:2000}. Furthermore,
imputing skewed outcomes under normality  leads to biased MI estimates of the distributional shape parameters \cite{hippel:2013}. 
We relax the normality assumption in MMRM by modeling the within subject dependence using the multivariate ST distribution\cite{azzalini:2003}, which 
includes the multivariate SN\cite{azzalini:1996}, normal and  t distributions as special cases. This extension is different from our previous work \cite{tang:2019},
where the data are modeled by a sequence of univariate ST regressions.

 The SN and ST distributions have a roughly bell-shaped density, and can be made arbitrarily close to the normal or t 
density by regulating suitable parameters \cite{azzalini:2003}. They are also capable of accommodate asymmetry and heavy tails often exhibited in clinical data \cite{liu:1995,jara:2008,lachos:2009}.
In the SN and ST distributions, the skewness is partially induced by truncating some  latent variables \cite{arnold:2002,valle:2006}.
Such  selection or truncation  mechanisms arise naturally in clinical studies. For example, a clinical trial may enroll only patients whose disease severity is above a certain level.  
In Section \ref{unimulst}, we briefly review the univariate and multivariate SN and ST distributions, and derive the relevant conditional distribution.

The SN and ST distributions present some undesirable properties in statistical inference \cite{liseo:2006, bayes:2007, branco:2013}.
This can be  illustrated in the univariate regression.  
In the scalar  case, the asymptotic distribution of the maximum likelihood estimate (MLE) is bimodal when the distribution of the data is close to normal, and there is a non-negligible chance that
the MLE of the skewness parameter can be infinite for skewed data \cite{pewsey:2000, Arellano:2008}. Similar issues exist in Bayesian inference \cite{ liseo:2006,  liseo:2013}.
In general,  the likelihood function changes slowly at large values of the skewness and/or  degrees of freedom (df)  parameters, and converges to a constant when the skewness   and/or df parameters
reach their limit values (all other parameters are  fixed). 
As a result,  the posterior distribution in Bayesian inference may be improper under   a diffuse prior, and
the posterior estimates  of the skewness and/or df parameters can be sensitive to the decay rate of their marginal prior densities  \cite{liu:1995, liseo:2006,Fonseca:2008}.  
The problems become  even more complex in the multivariate case.
Section \ref{priorst} investigates how to specify the priors for the skewness, df and covariance parameters to address the inference challenges mentioned above.

In Sections \ref{mdanopattern} and \ref{mdapattern}, we extend Tang's  MDA algorithm \cite{2017:tangb,2017:tanga} to the multivariate ST, SN and t regressions. The underlying idea is to reorganize  these robust regressions as the normal linear regression with the introduction of some latent variables.
The parameter expansion (PX \cite{liu:1999, liu:2000})  and  block sampling techniques are employed to improve the mixing and accelerate the convergence of the Markov chain.
Section \ref{impdrop} applies the MDA algorithm to the controlled imputation.
The proposed MDA algorithm and missing data imputation methods are illustrated by real data analyses in Section \ref{numexam}.

Throughout the article, the following notations will be used. Let $\mathcal{G}(a,b)$ denote a gamma distribution with shape  $a$, rate $b$ and mean $a/b$.
Let  $N(\muv,\Omega)$ denote  the  normal distribution with mean $\muv$ and covariance $\Omega$, and $t(\muv,\Omega,\nu)$ the  $t$ distribution  with  $\nu$ df. 
In the multivariate case, $\muv$ is a vector and $\Omega$ is a square matrix.
The probability  density functions (PDF) of the gamma, normal and t distributions are denoted, respectively, by
$\mathcal{G}(\cdot|\cdot)$, $N(\cdot|\cdot)$ and $t(\cdot|\cdot)$. 
 Let $\Phi(\cdot)$ and $T_\nu(\cdot)$ denote respectively the cumulative distribution function (CDF) of $N(0,1)$ and $t(0,1,\nu)$.
Let $N^+(\mu,\sigma^2)$ be the scalar positive normal distribution  left truncated at $0$,  and $t^+(\mu,\sigma^2,\nu)$ the scalar positive t distribution.
Let $\text{W}^{-1}(A, n_0)$ denote the inverse Wishart distribution with  PDF $\pi(\Sigma)\propto |A|^{n_0/2}|\Sigma|^{-(n_0+p+1)/2}\exp[-\text{tr}(A\Sigma^{-1})/2]$.

\section{Review of univariate and multivariate SN and ST distributions}\label{unimulst}
We  review the univariate and multivariate SN and ST distributions introduced by Azzalini and his collaborators \cite{AZZALINI:1985,azzalini:1996,azzalini:2003}. We focus on
 their convolution-type stochastic representation \cite{azzalini:2003,lee:2013a} as
it allows straightforward interpretation of the skewness parameters  in the multivariate distribution (see Section  \ref{skewmult} below), and
makes it easier to  design the MCMC algorithm \cite{schnatter:2010}. 
An alternative stochastic representation is given by Azzalini and Capitanio \cite{azzalini:2003}.
In both representations, the skewness is induced  by truncating a latent variable \cite{arnold:2002}. Such mechanism arises naturally in practice.
 For example, in clinical trials, patients may be selected only if a variable of interest is above a threshold. 

\subsection{Univariate SN and ST distributions}\label{skewuni}
The PDF of a SN \cite{AZZALINI:1985} random variable $y\sim \mathcal{SN}(\mu,\sigma^2,\psi)$   is given by $2N(y|\mu,\omega^2)\Phi[\lambda (y-u)/\omega]$, where  $\omega^2=\sigma^2+\psi^2$. It can be stochastically represented as
$$y = \mu + \psi \wc+ \epsilon, $$
where $\mu$ is the location parameter, $\text{\small$\mathcal{W}$} \sim N^+(0,1)$ is independent of  $\epsilon\sim N(0,\sigma^2)$, and  $\lambda=\psi/\sigma$ is the skewness parameter.
Its mean is $\text{E}(y)=\mu+\psi\sqrt{\frac{2}{\pi}}$.

The parameter $\lambda$ controls the degree to which the data depart from normality.
 When $\lambda=0$, the SN distribution reduces to the normal distribution.
The degree of the skewness of $y$ increases as the absolute value of $\lambda$ increases. 
Since the maximum  skewness and kurtosis are  $0.995$ and  $0.869$ respectively  \cite{AZZALINI:1985}, 
the SN distribution is suitable only for mildly or moderately non-normal data. 

The ST distribution\cite{azzalini:2003}  allows a higher degree of skewness and  kurtosis. A ST random variable  $y\sim \mathcal{ST}(\mu,\sigma^2,\psi,\nu)$
can be stochastically represented as
\begin{equation}\label{skewtdist}
y = \mu + \frac{1}{\sqrt{d}}  [\psi \wc^* + \epsilon] =  \mu + \psi \wc + \frac{1}{\sqrt{d}}\epsilon,
\end{equation}
where $d \sim \chi_\nu^2/\nu$ [i.e. $d\sim \mathcal{G}(\nu/2,\nu/2)$], $\wc^* \sim N^+(0,1)$,  $\wc=\wc^*/\sqrt{d}\sim t^+(0,1,\nu)$ and $\epsilon\sim N(0,\sigma^2)$. We get $\text{E}(y)=\mu+\psi\sqrt{\frac{\nu}{\pi}} \frac{\Gamma((\nu-1)/2)}{\Gamma(\nu/2)}$, where $\Gamma(\cdot)$ is the gamma function. 
 The PDF of $y$ is given by
\begin{equation}\label{distat2}
f_{_\mathcal{ST}}(y| \mu, \omega^2, \lambda,\nu)=2 t (y|\mu,\omega^2,\nu) \,T_{\nu+1} \left[\lambda \frac{y-\mu}{\omega}\sqrt{\frac{\nu+p}{\nu+\frac{(y-\mu)^2}{\omega^2}}}\right].
\end{equation}

\subsection{Multivariate SN and ST distributions}\label{skewmult}
A  multivariate version of the SN distribution is introduced by Azzalini  and Dalla Valle \cite{azzalini:1996}. The random vector $\yv=(y_1,\ldots,y_p)' \sim \mathcal{SN}(\muv, \Sigma, \psiv)$ 
can be represented as \cite{schnatter:2010}
\begin{equation}\label{dista1}
(y_{1},\ldots, y_{p})' = \muv + \psiv \wc+ \epsilonv
\end{equation}
where $\epsilonv= (\epsilon_{1},\ldots, \epsilon_{p})' \sim N(\zerov, \Sigma)$, $\wc \sim N^+(0,1)$, $\muv=(\mu_1,\ldots,\mu_p)'$ is a vector containing the location parameters, and
$\psiv=(\psi_1,\ldots,\psi_p)'$ is a vector of skewness parameters. 
The PDF of $\yv$ is given by
\begin{equation}\label{dista2}
f_{_\mathcal{SN}}(\yv| \muv,\Sigma, \psiv)= 2N (\yv|\muv,\Omega)\Phi[\lambdav^{*'} (\yv-\muv)]= 2N (\yv|\muv,\Omega)\Phi\left[\sum_{j=1}^p \lambda_j \frac{y_j-\mu_j}{\omega_j}\right] ,
\end{equation}
where  $\omega_j^2$ is the $(j,j)$th element of $\Omega= \Sigma +\psiv \psiv'$, and
$\lambdav^* =\left(\frac{\lambda_1}{\omega_1},\ldots,\frac{\lambda_p}{\omega_p}\right)= \frac{\Omega^{-1}\psiv}{ \sqrt{1-\psiv' \Omega^{-1}\psiv}}= \frac{\Sigma^{-1}\psiv}{  \sqrt{1+\psiv' \Sigma^{-1}\psiv}}$.

The multivariate SN distribution \cite{azzalini:1996} was originally introduced  through the parametrization $(\muv,\lambda_1,\ldots,\lambda_p,\Omega)$. It is difficult to interpret
the skewness parameter $\lambda_j$, which does not provide information about the skewness of $y_j$, not even on its sign \cite{Arellano:2008}. Also, $\Omega$ is not the
covariance matrix of $\yv$. Let $\Sigma_{jj}=\text{var}(\epsilon_j)$. By the stochastic representation \eqref{dista1}, the marginal distribution of $y_j= \mu_j +\psi_j\wc +\epsilon_j$ is 
$SN(\mu_j, \Sigma_{jj},\psi_j)$, and its 
 skewness  is controlled by $\psi_j/\sqrt{ \Sigma_{jj}}$.

The multivariate ST distribution \cite{azzalini:2003} $\yv \sim \mathcal{ST}(\muv, \Sigma, \psiv,\nu)$
can be represented as
\begin{equation}\label{skewtdist}
 (y_{1},\ldots, y_{p})' = \muv + \psiv \wc +\frac{1}{\sqrt{d}} \epsilonv,
\end{equation}
where $d \sim \chi_\nu^2/\nu$, $\wc^*\sim N^+(0,1)$ and  $\wc=\wc^*/\sqrt{d}\sim t^+(0,1,\nu)$. 
The PDF of $\yv$ is 
\begin{equation}\label{distat2}
f_{_\mathcal{ST}}(\yv|\muv, \Sigma, \psiv,\nu)= 2t (\yv|\muv,\Omega,\nu) T_{\nu+p} \left[\lambdav^{*'} (\yv-\muv)\sqrt{\frac{\nu+p}{\nu+(\yv-\muv)'\Omega^{-1}(\yv-\muv)}}\,\right].
\end{equation}

Let the LDL decomposition of $\Sigma$ be denoted by  $\Sigma = L \Lambda L'$,
where $\Lambda=\text{diag}(\gamma_1^{-1},\ldots,\gamma_p^{-1})$, $L=U^{-1}$, and $U=$ {\small$\begin{bmatrix} 1 & 0 &  \ldots & 0\\
                       -\beta_{21} & 1 & \ldots &0 \\
                       
                      -\beta_{p1} & \ldots & -\beta_{p,p-1} & 1 \\
                    \end{bmatrix}$}. 
Equation \eqref{skewtdist} can be reorganized as $U\yv=U\muv + U\psiv \wc +\frac{1}{\sqrt{d}}  U\epsilonv$ or equivalently as
\begin{equation}\label{skewdistseq0}
 y_{j} =\sum_{t=1}^{j-1} \beta_{jt} y_{t} +\underline \mu_j +\underline \psi_j \wc+\frac{\varepsilon_j}{\sqrt{d}} \text{ for } j=1,\ldots,p,
\end{equation}
where  $(\underline{\mu}_j,\underline{\psi}_j )=(\mu_j,\psi_j )-\sum_{t=1}^{j-1}\beta_{jt}(\mu_t,\psi_t)$, and  $\varepsilon_j$'s  are independently distributed as $\varepsilon_j \sim N(0,\gamma_j^{-1})$.

 It is well known \cite{azzalini:1996,azzalini:2003} that the conditional distribution of  $\yv_{s_2}=(y_{s+1},\ldots,y_p)'$ given  $\yv_{s_1}=(y_1,\ldots,y_s)'$ is
 no longer the SN/ST distribution.
As shown in Lemma \ref{condAversion} below, the conditional distribution of $\yv_{s_2}$ given  $\yv_{s_1}$ can be expressed similarly 
to Equation \eqref{skewdistseq0}   except that the location parameter for the positive normal or t random variable  is not $0$.
We omit the proof since a more general result is given in Appendix \ref{postwdint}.

\begin{lemma}\label{condAversion}
(a) The conditional distribution of $(d, \wc)$ given $\yv_{s_1}=(y_1,\ldots, y_s)'$ is 
\begin{eqnarray}\label{postdw1}
\begin{aligned}
\wc | y_1,\ldots,y_s  \sim t^+\left[\frac{B}{A}, \frac{\nu_{d}}{A(\nu+s)},\nu+s\right] \,\text{ and }\, \\
d|\wc,y_1,\ldots,y_s  \sim \mathcal{G}\left[\frac{\nu+s+1}{2}, \frac{\nu_{d}+ A(\wc-\frac{B}{A})^2}{2}\right],
\end{aligned}
\end{eqnarray}
where $y_j^*=y_j-\sum_{t=1}^{j-1}\beta_{jt}y_t-\underline{\mu}_j$, 
$A=1+\sum_{j=1}^s \gamma_j\underline\psi_j^2$,
$B=\sum_{j=1}^s \gamma_j \underline\psi_j y_j^*$ and $\nu_d=\nu+\sum_{j=1}^s \gamma_j y_j^{*^2} -B^2/A$. \\
(b)  
 The conditional distribution of $\yv_{s_2}$ given $\yv_{s_1}$ can be represented as a sequence of univariate conditional distributions
$$ y_{j}= \sum_{t=1}^{j-1}\beta_{jt}y_t + \underline\mu_{j}+ \underline\psi_{j} \wc_s +\frac{1}{\sqrt{d_s}} \varepsilon_j\, \text{ for }\, j\geq s+1,$$
where $\varepsilon_{k} \sim N(0,\gamma_k^{-1})$, $\varepsilon_k$'s are independent, and $(\wc_s, d_s)$ is the random sample from the conditional distribution \eqref{postdw1}.\\
(c) The conditional distribution of $\yv_{s_2}$ given $\yv_{s_1}$ can be equivalently   represented in matrix form as 
\begin{eqnarray*}
\begin{aligned}
\begin{bmatrix} y_{s+1} \\ \ldots \\ y_p\end{bmatrix} 
=  \begin{bmatrix} \mu_{s+1} \\ \ldots \\ \mu_p \end{bmatrix}  -U_{s_{22}}^{-1} U_{s_{12}} \begin{bmatrix} y_{1}-\mu_1 \\ \ldots \\ y_s-\mu_s \end{bmatrix}  +U_{s_{22}}^{-1}  \begin{bmatrix} \underline\psi_{s+1} \\ \ldots \\ \underline\psi_p \end{bmatrix}\wc_s
  + \frac{U_{s_{22}}^{-1}}{\sqrt{d_s}}  \begin{bmatrix} \varepsilon_{s+1}\\ \ldots \\ \varepsilon_p \end{bmatrix}, 
\end{aligned}
\end{eqnarray*}
where $\begin{bmatrix} U_{s_{11}} & \zerov \\
                                        U_{s_{21}} & U_{s_{22}} \\\end{bmatrix} =   $
{\small $\left[\begin{array}{ccc|cc} 1 & 0 & 0 & \ldots & 0\\
                              \ldots & & & & \\
                       -\beta_{s1} & \ldots & 1 & \ldots &0 \\\hline
                            \ldots & & \ldots & \ldots & 0 \\         
                      -\beta_{p1} & \ldots & -\beta_{ps} &  \ldots & 1 \\
                    \end{array}\right]$} is a partition of $U$.
\end{lemma}

{\flushleft{\bf Notes:}}\\
1. In Lemma  \ref{condAversion}(a), the conditional distribution of $d$ given $y_1,\ldots,y_s$  is not gamma due to the constraint $\wc>0$.\\
2. Lemma \ref{condAversion} can be applied to the multivariate SN distribution by setting $\nu\equiv\infty$, $d\equiv 1$ and $\wc| \yv_{s_1} \sim N^{+}(\frac{B}{A}, \frac{1}{A})$.\\
3. In the multivariate t distribution ($\psi_j\equiv 0$), the conditional distribution of $\yv_{s_2}$ given $\yv_{s_1}$ can be written as
$y_{j} = \sum_{t=1}^{j-1} \beta_{jt} y_t + \underline\mu_{j} + \frac{\varepsilon_j}{\sqrt{d_s}}$ for $j>s$, where 
$d_s \sim  \frac{\chi_{\nu+s}^2}{\nu+ \sum_{t=1}^s \gamma_t y_t^{*^2}}$.\\
4. The skewness of the conditional distribution of $y_{s+1}$ given $y_1,\ldots,y_{s}$ is a monotone function of  $\underline\psi_j \sqrt{\gamma_j}$.

\section{MDA algorithm for MMRM with no restriction on fixed effects}\label{mdascheme}
Suppose a study consists of $n_{tot}$ subjects, and the data are collected at $p$ fixed time points.  Let $\yv_i =(y_{i1},\ldots,y_{ip})'$ denote
the complete outcome  for subject $i$. 
In general,  $\yv_i$'s won't be fully observed.
Let $s_i$ be the last visit that  subject $i$ has a measurement observed, and $s_i=0$ if the subject has no observed outcome. 
Let $\yv_{io}$, $\yv_{im}$ and $\yv_{iw}$ denote respectively the observed data with $o_i$ elements, intermittent missing data with $m_i=s_i-o_i$ elements prior to dropout, and
missing data after dropout for subject $i$. 
Without loss of generality, we sort the data  so that subjects in pattern $s$ are arranged before subjects in pattern $t$ if $s>t$.
Let $n_j$ be the total number of subjects in patterns $j,\ldots,p$. Let $n=n_1$ be the number of subjects with at least one observed outcome.

The following MMRM is often used to analyze longitudinal outcomes collected at a  number of fixed time points \cite{aiddiqui:2009, laird:1987}
\begin{equation}\label{mmrm_1}
 y_{ij}= \sum_{k=1}^{q} \alpha_{kj} x_{ik}  + \xi_{ij} \text{ for } j=1,\ldots,p,
\end{equation}
 where $i=1,\ldots,n_{tot}$ indexes subjects, $q$ is the number of baseline covariates, 
 $\alpha_{kj}$ is the effect of covariate $x_{ik}$  at visit $j$, and
 $\xiv_i=(\xi_{i1}, \ldots,\xi_{ip})' \sim N(0,\Omega)$.
Let $x_{i1}\equiv 1$ if the model contains an intercept term. In clinical trials, we set the treatment status as
 $x_{iq}=g_i=1$ for  the experimental treatment, and $g_i=0$ for the control treatment.  
We place no constraints on   the covariate effects  $\alpha_{kj}$'s, and they can vary freely over time.
As discussed in Section \ref{intsec}, we employ an unstructured covariance matrix. 
It generally provides a good control of the type I error rate under the null hypothesis, and
 results in a negligible loss in efficiency\cite{lu:2010}  under the alternative hypothesis when compared to the analysis based on the true  covariance structure (it is difficult to be prespecified) in moderate to large samples. 
A structured covariance matrix, which may be induced through the use of random effects, is useful when individuals have  a large number of observations, or varying  times of observations \cite{laird:1987}.

Model \eqref{mmrm_1} assumes that the outcomes are normally distributed. Inference about the treatment effect based on the normality assumption  can be inefficient in the presence of outliers \cite{pinheiro:2001,zhang:2001}, and may be invalid for highly non-normal data particularly
when the sample size is small \cite{arnau:2013}.
In this article, we use the multivariate t, SN or ST distribution to model skewed and/or fat-tailed data. 
Model \eqref{mmrm_1} will be denoted  respectively by MMRM-n, MMRM-t, MMRM-sn or MMRM-st when the residual errors  $\xiv_i=(\xi_{i1}, \ldots,\xi_{ip})'$ are modeled by the multivariate 
normal, t, SN or ST distribution. 

By introducing the latent variables $(d_i,\wc_i)$, we can expand the MMRM-st  as
\begin{equation}\label{mixed}
y_{ij}=   \sum_{k=1}^{q} \alpha_{kj} x_{ik}   +  \psi_{j} \wc_i + \frac{1}{\sqrt{d_i}}  \epsilon_{ij}=  \sum_{k=1}^{\Qs} \alpha_{kj} x_{ik} +\frac{1}{\sqrt{d_i}}  \epsilon_{ij},
\end{equation} 
where  $d_i \sim \mathcal{G}(\nu/2,\nu/2)$, $\wc_i^*\sim N^+(0,1)$, $\wc_i=\wc_i^*/\sqrt{d_i}$, 
 $\Qss=q+1$,   $\alpha_{\Qs j}=\psi_j$, and   $x_{i\Qs}=\wc_i$. 
The MMRM-sn and MMRM-t can be obtained by
setting $d_i\equiv 1$ (i.e. $\nu\equiv\infty$) and $\psi_j\equiv 0$ respectively in model \eqref{mixed}. 

Model  \eqref{mixed}  can be reorganized as the  product of the following conditional  models
\begin{eqnarray}\label{factor} 
y_{ij} \sim N\left( \sum_{k=1}^{\Qs} \underline\alpha_{kj} x_{ik}+ \sum_{t=1}^{j-1}\beta_{jt}y_{it}, (d_i\gamma_j)^{-1}\right)
\text{ for } j=1,\ldots,p,
\end{eqnarray}
 where    $\underline{\alpha}_{kj} =  \alpha_{kj} - \sum_{t=1}^{j-1} \beta_{jt} \alpha_{kt}$.
Note that   $\underline\alpha_{\Qs j} =\underline\psi_j=\psi_j - \sum_{t=1}^{j-1} \beta_{jt} \psi_t$.

\subsection{The prior distribution}\label{priorst}
We assume that   $\nu$ and $(\Sigma,\alphav, \underline\psi_1,\ldots,\underline\psi_p)$ are independent, and that $\alphav, \underline\psi_1,\ldots,\underline\psi_p$ are conditionally independent given $\Sigma$ in the prior distribution,
 where  $\alphav=\begin{bmatrix} \alpha_{11} & \ldots & \alpha_{q1} \\
                                                               & \ldots & \\
                                                          \alpha_{1p} & \ldots &\alpha_{qp} \end{bmatrix}$.
We use  noninformative or objective priors  in our  numerical examples,  but our  specification allows informative priors. 
The missing data imputation is based on  model \eqref{factor} with the parameterization  $\{ \nu, (\thetav_1,\gamma_1),\ldots,(\thetav_p,\gamma_p)\}$,
where  $\thetav_j=(\underline\alpha_{1j},\ldots,\underline\alpha_{\Qs j},\beta_{j1},\ldots,\beta_{jj-1})'$. The prior on $(\thetav_j,\gamma_j)$'s can be induced from the prior
on $(\Sigma,\alphav, \underline\psi_1,\ldots,\underline\psi_p)$.  It is also possible to place priors directly on $(\thetav_j,\gamma_j)$'s.

\subsubsection{Prior on $\Sigma$} We employ the hierarchical prior introduced by Huang and Wand \cite{huang:2013}
$$  \rho_j\stackrel{i.i.d}{\sim} \mathcal{G}\left(\frac{1}{2},\frac{1}{a_0^2}\right) \text{ and } \Sigma|\rho_1,\ldots,\rho_p \sim \text{W}^{-1}( A_w,n_w),$$
where $n_w= n_0+p-1$, $A_w= 2n_0\,\text{diag}(\rho_1,\ldots,\rho_p)$. 
It is an extension of the  half-t prior  \cite{gelman:2006} used in a hierarchical model of
variance parameters.
The choice of $n_0=2$ and a large value for $a_0$ (e.g. $a_0=10^5$) corresponds to highly noninformative half-t priors on each standard deviation term  and uniform priors on each correlation term\cite{huang:2013}.

The inverse Wishart distribution $\text{W}^{-1}(A_w, n_w)$ with fixed $A_w$ and $n_w$ is commonly used as the prior for $\Sigma$ in the multivariate normal and t linear regressions \cite{liu:1995, schafer:1997, 2015a:tang}. It reduces to  Jeffrey's prior at $n_w=0$ and $A_w=\zerov$.  
The  inverse Wishart or Jeffrey's prior can be quite informative or inappropriate in the multivariate SN and ST regression for  highly skewed data, and the argument is the same as that in the univariate regression \cite{tang:2019}.

\subsubsection{Prior on the skewness parameters}
In the scalar case, the Bayes estimate of the skewness parameter $\lambda=\psi\sqrt{\gamma}$ can be infinite under a diffuse prior since there is a non-negligible 
chance that the likelihood function is a monotone function of $\lambda$ at fixed $(\mu,\omega^2)$. 
The issue can be resolved by using the objective prior for $\lambda$ \cite{liseo:2006, bayes:2007}. The prior has no closed-form  expression, 
but can be well approximated \cite{liseo:2006, branco:2013} by  $\lambda\sim t(0,\pi^2/4,1/2)$, or equivalently by $\psi|\gamma \sim t(0,\pi^2/4\gamma,1/2)$. 
No appropriate objective prior has been developed  in  the multivariate case \cite{branco:2013}.
We set the prior $\underline\psi_j|\Sigma \sim t(0,\pi^2/4\gamma_j,1/2)$ for the skewness parameter $\underline\psi_j$ of $\underline{y}_{ij}$, and
 assume  $\underline\psi_j$'s   are conditionally independent given $\Sigma$ in the prior.  
The prior is equivalent to the following hierarchical prior 
$$ d_{j_\ps} \sim \mathcal{G}\left(\frac{1}{4},\frac{1}{4}\right)  \text{ and } \underline\psi_j|\Sigma,d_{j_\ps} \sim N\left(0, \frac{\pi^2}{4d_{j_\ps}\gamma_j}\right)  \text{ for } j=1,\ldots,p.$$

Liseo and Parisi \cite{liseo:2013} specifies  a prior that  requires
certain constraints on the skewness parameters $\psi_j$'s to make $\Omega$ positive definite \cite{liseo:2013}, and assumes conditional independence among 
the skewness parameters $\psi_j$'s for $y_{ij}$'s  given $\Sigma$. 
Our prior  is more convenient to use, and might be  more reasonable since $\underline{y}_{ij}$'s tend to be less correlated than $y_{ij}$'s. 

\subsubsection{Prior on  \texorpdfstring{$\alphav$}{\alpha}}
Suppose the prior for $\alphav$  is 
$$\text{vec}(\alphav)|\Sigma \sim N(\text{vec}(\alphav_0), M^+\otimes \Sigma),$$
where $M$ is a given $q \times q$  covariance matrix with rank $r^*$, and $M^+$ is the Moore-Penrose inverse of $M$.
We allow $M$ to be degenerate \cite{2015a:tang}. If all elements in the $k$-th column of $\alphav_0$, and  
all elements in the $k$-th row and $k$-th column of $M$ are  $0$, the prior for the effects $(\alpha_{k1},\ldots,\alpha_{kp})$ of covariate $x_{ik}$  is flat.
We set  $M=\zerov$ and $\alphav_0=\zerov$ in our  examples.

\subsubsection{Prior on $($\texorpdfstring{$\thetav$}{\theta}$_j,\gamma_j)$'s}
 The prior on $(\thetav_1,\gamma_1,\ldots,\thetav_p,\gamma_p)$ can be induced from the prior on $(\alphav,\underline\psi_1,\ldots,\underline\psi_p, \Sigma)$. By Lemma $2$ of Tang \cite{2015a:tang}, we get 
\begin{eqnarray}\label{priorgen}
\begin{aligned}
 d_{j_\ps}\sim \mathcal{G}(1/4,1/4) \text{ and } \pi(\theta_j, \gamma_j) \propto \gamma_j^{\frac{n_w+2j+r-p-1}{2}-1} \exp\left[-\frac{\gamma_j}{2} \tilde\thetav_j' E_j \tilde\thetav_j\right] 
\end{aligned}
\end{eqnarray}
for $j=1,\ldots,p$,
where $\tilde\thetav_j=(-\thetav_j',1)'$, $E=  \begin{bmatrix}  M & \zerov & M\alphav_0'  \\
                                                                                                                                \zerov' & \frac{ 4d_{j_\ps}  }{\pi^2} & \zerov' \\
                                                                                                                     \alphav_0 M & \zerov & \alphav_0 M \alphav_0' +A_w\\ \end{bmatrix} $ is  a $(\Qss+p)\times (\Qss+p)$  matrix,  $E_j$
is the $(\Qss+j)\times (\Qss+j)$ leading principle submatrix of $E$,  and $r=r^*+1$ is the rank of  the matrix $\text{diag}(M,\frac{ 4d_{j_\ps}  }{\pi^2})$. Since $\alphav_0=\zerov$ and $M=\zerov$, we get $r^*=0$ and $r=1$.
 We do not recommend specifying $M$ by a diagonal matrix  with small diagonal elements  because   $\gamma_j$'s tend to be  overestimated particularly in small samples
since
the shape parameter of the posterior gamma distribution of $\gamma_j$ will increase by $r^*/2=q/2$,  but the rate parameter changes little.

\subsubsection{Prior on $\nu$}
In the Student's t regression, the  likelihood function converges to a constant as $\nu \rightarrow \infty$   (other model parameters are fixed) since the t distribution converges to the Gaussian distribution. 
As a result, the posterior distribution
of $\nu$ is proper only if the decay rate of the prior density of $\nu$ satisfies certain conditions, and
inference on $\nu$ is quite sensitive to the shape of  the prior density of $\nu$ \cite{liu:1995,Fonseca:2008}.  The same issue exists for the ST regression because  the
ST distribution converges to the SN distribution as $\nu\rightarrow \infty$.

We use the penalized complexity (PC) prior \cite{simpson:2017}   since  it shows good performance in the  Student's t regression in simulation. It  is obtained
by penalizing the complexity between the multivariate t distribution $t(\mu,\frac{\nu-2}{\nu}\Sigma,\nu)$ and the normal distribution $N(\mu,\Sigma)$. 
The density of the PC prior  is determined numerically in Simpson {\it et al}  \cite{simpson:2017}. Its analytic expression is derived 
in Appendix \ref{prepostnu}
  $$\pi(\nu){{=}} \lambda \exp(-\lambda d(\nu)) \left|\frac{\partial d(\nu)}{\partial{\nu}}\right|,$$
where  $\Psi(\cdot)$ and $\Psi'(\cdot)$ are  the  digamma and trigamma functions, $b(\nu)=\Psi(\frac{\nu+p}{2})-\Psi(\frac{\nu}{2})$,
$$d(\nu)= \sqrt{p\left[1+\log\left(\frac{2}{\nu-2}\right)\right] +2\log\frac{\Gamma (\frac{\nu+p}{2})}{\Gamma (\frac{\nu}{2})}-(\nu+p)b(\nu)} \,\text{ and }\,$$
$$\frac{\partial d(\nu)}{\partial{\nu}}= -\frac{\frac{p}{\nu-2} +\frac{\nu+p}{2}[\Psi'(\frac{\nu+p}{2})-\Psi'(\frac{\nu}{2})] }{{{ 2}}d(\nu)}.$$
{ Tang\cite{tang:2019} gives the analytic  density when $p=1$, where $\frac{\partial d(\nu)}{\partial{\nu}}$ is  wrongly written as $\frac{\ldots}{4d(\nu)}$ (detected by Dr Rue), but it does not affect the MCMC sampling and  inference results.}
In the PC prior, $\nu$ is bounded below by $\nu_l=2$.
We also put an upper bound $\nu_m=1000$ on $\nu$ since the  PDF can not be accurately calculated due to  rounding errors at  large values of $\nu$.
Another popular prior for $\nu$ is the reference prior given by Fonseca {et al} \cite{Fonseca:2008}.

\subsection{MDA algorithm}\label{mdanopattern}
The joint posterior distribution of $\phiv=\{ (\thetav_1,\gamma_1,d_{1_\ps},\rho_1),\ldots, (\thetav_p,\gamma_p,d_{p_\ps},\rho_p)\}$,  $\nu$ and $(d_i,\wc_i, \yv_{im})$'s is given by 
\begin{eqnarray}\label{jointpos}
\begin{aligned}
&\pi(Y_m, d_i\text{'s},\wc_i\text{'s},\phiv,\nu |Y_o) \\
& \,\,\propto 
\pi(\nu)\pi(\phiv)   \left[\prod_{i=1}^{n} f(d_i, \wc_i|\nu)\prod_{j=1}^p\prod_{i=1}^{n_i} f(y_{ij}|y_{i1},\ldots,y_{ij-1},d_i,\wc_i,\phiv,\nu)\right],
\end{aligned}
\end{eqnarray}
where $Y_{o}=\{\yv_{io}: i=1,\ldots,n\}$, $Y_{m}=\{\yv_{im}: i=1,\ldots,n\}$, the last term in the bracket is the likelihood for the augmented data $(d_i\text{'s},\wc_i\text{'s}, Y_o,Y_m)$, and
 $$f(d_i, \wc_i|\nu)\propto  d_i^{\frac{\nu+1}{2}-1} \exp[-d_i \frac{\nu+\wc_i^2}{2}],$$
$$f(y_{ij}|y_{i1},\ldots,y_{ij-1},d_i,\wc_i,\phiv,\nu) \propto \sqrt{d_i\gamma_j} \exp\left[-\,\frac{d_i\gamma_j(y_{ij} - \sum_{k=1}^\Qs \underline\alpha_{kj} x_{ik}- \sum_{t=1}^{j-1}\beta_{jt}y_{it})^2}{2}\right],$$
$$\pi(\phiv) \propto \prod_{j=1}^p \left\{\rho_j^{\frac{1+n_w}{2}-1}\exp(-\frac{\rho_j}{a_0^2}) d_{j_\ps}^{\frac{1}{4}-1}\exp(-\frac{d_{j_\ps} }{4})
  \gamma_j^{\frac{n_w+2j+r-p-1}{2}-1} \exp\left[-\frac{\gamma_j}{2} \tilde\thetav_j' E_j \tilde\thetav_j\right] \right\}.$$

 In the MDA algorithm, the missing data $\yv_{iw}$ after dropout are integrated out of the posterior distribution, and imputed after the algorithm converges. The details will be given in Section \ref{impdrop}.
Subjects without any observed outcomes (i.e. in pattern $0$) will not be used in the posterior sampling of the model parameters and $(\yv_{im}, d_i,\wc_i)$'s.

The MDA algorithm for MMRM-st  repeats the following steps until convergence
\begin{itemize}
\item[P0.] Update $\rho_j$'s from its posterior distribution $\mathcal{G}(\frac{n_0+p}{2}, n_0 \sum_{k=j}^p\gamma_k\beta_{kj}^2+\frac{1}{a_0^2})$ 
\item[P1.] Update $(d_{\psi_j}, \theta_j, \gamma_j)$'s by  drawing  $d_{\psi_j}$ from $\mathcal{G}(\frac{3}{4},\frac{1}{4}+\frac{2}{\pi^2} \gamma_j\underline\psi_j^2)$, and 
$(\theta_j, \gamma_j)$  from the gamma-normal posterior distribution using Tang's  method \cite{2015a:tang, tang:2019}
\begin{equation}\label{postgamma}
 \pi(\thetav_j,\gamma_j|\phiv,\nu, Y_m,Y_{o})  \propto \gamma_j^{ \frac{n_j+n_w+r+2j-p-3}{2}}\exp\left[-\frac{ \gamma_j  \tilde\thetav_j'(\sum_{i\leq n_j} d_i\tilde{\xv}_{ij}\tilde{\xv}_{ij}'+E_j) \tilde\thetav_j  }{2}\right],
\end{equation}
where  $\tilde{\xv}_{ij}=(x_{i1},\ldots, x_{i\Qs}, y_{i1}, \ldots, y_{ij})'$. 
\item[P2.] Update $\nu$ by a MH sampler with the proposal distribution  $\log(\nu^*-\nu_l)\sim N[\log(\nu-\nu_l), c^2]$.
The  parameter $c$ is tuned to make the acceptance probability  lie roughly in the range of $30-70\%$.  The details are given in Appendix \ref{prepostnu}.
\item[I.] Impute $(d_i, \wc_i, \yv_{im})$ given $(\phiv, \nu,Y_o)$  from their posterior distribution \eqref{postdwy} given in Appendix \ref{postwdint}.
\item[PX1.]  Update $(d_1,\ldots,d_n, \gamma_1,\ldots,\gamma_p)$ as $(gd_1,\ldots,gd_n, \gamma_1/g,\ldots,\gamma_p/g)$, where $g$ is  drawn from its posterior distribution given in  Appendix \ref{postg}.
\item[PX2.] Update $(\wc_1,\ldots,\wc_n,\underline\psi_1,\ldots,\underline\psi_p) \rightarrow (h \wc_1,\ldots,h \wc_n,\underline\psi_1/h,\ldots,\underline\psi_p/h)$, where
$H=h^2$ is drawn from its posterior distribution given in  Appendix \ref{posth}.
\end{itemize}

{\flushleft{\bf Notes:}}\\
1. Steps P2 and I form a block $(\nu, \wc_i\text{'s},d_i\text{'s},\yv_{im}\text{'s})$ in the sense that they are drawn from 
\begin{eqnarray*}
\begin{aligned}
& \pi(\nu, \wc_i\text{'s},d_i\text{'s},\yv_{im}\text{'s} | \phiv,Y_o ) \\
& \,\, \propto   \pi(\nu| \phiv,Y_o) \prod_{i=1}^{n}\left\{\pi( \wc_i |\nu, \phiv,Y_o) \pi(d_i| \wc_i,\nu,\phiv,Y_o) \pi(\yv_{im} | d_i,\wc_i,\nu,\phiv,Y_o)\right\}.
\end{aligned}
\end{eqnarray*}
The use of the blocking technique generally reduces the autocorrelation between posterior samples and speeds up the convergence of the Markov chain.

In the above approach, sampling $\nu$ requires calculating the marginal density of $\yv_{io}$. 
As described in Appendix \ref{prepostnu},  the  density of $\yv_{io}$ can be  computed without matrix inversion for monotone missing data.
Therefore, one alternative strategy is to replace steps P2 and I by sampling $(\nu, \wc_i\text{'s},d_i\text{'s})$ from their posterior distribution  $\pi(\nu, \wc_i\text{'s},d_i\text{'s} | \phiv, Y_o,Y_m)$ 
using the method described  in Appendix  \ref{prepostnu} (applied after the intermittent missing data $\yv_{im}$'s are filled in), and 
imputing  $\yv_{im}$'s  from the posterior distribution  $\pi(\yv_{im}| d_i, \wc_i,\yv_{io},\phiv,\nu)$  given in Equation \eqref{postym} in Appendix \ref{postwdint}.  

Another strategy is to keep step I unchanged, but update $\nu$   from its conditional posterior distribution given  $(\phiv, Y_o,Y_m,  \wc_i\text{'s},d_i\text{'s})$  via the MH sampler in step P2 
  $$\pi(\nu|\phiv,Y_o,Y_m,\wc_i\text{'s},d_i\text{'s}) \propto \pi(\nu)  \frac{(\nu/2)^{n\nu/2}}{\Gamma^n(\nu/2)} \left[\prod_{i=1}^n d_i\right]^{\nu/2-1}\exp\left[-\frac{\nu\sum_i d_i}{2}\right] I(\nu>\nu_l).$$
The per step computational time is reduced, but it may take many more iterations for the algorithm to converge with larger  autocorrelation between the posterior samples of $\nu$.\\
2.  Steps PX1 and PX2 are the generalized Gibbs samplers \cite{liu:1999,liu:2000} used to accelerate the  convergence of the algorithm. Omitting the two steps does not alter the  stationary distribution of the Markov chain.
 Inclusion of these steps tends to improve the  mixing of the chain.  \\
3.   In step P1, we can also draw $(\theta_{j}, \gamma_j)$  via the MH sampler based directly on the Student's t prior for $\underline\psi_j$'s ($d_{\psi_j}$ is integrated out  of the prior).
 The candidate $(\thetav_j^*,\gamma_j^*)$ is drawn from $ \gamma_j^{\frac{n_j+n_0+2j+(r^*+1)-p-3}{2}} \exp\left[-\frac{\gamma_j}{2} \tilde\thetav_j'(E_j+\sum_{i\leq n_j} d_i\tilde{\xv}_{ij}\tilde{\xv}_{ij}')\tilde\thetav_j\right]$, and accepted with probability $\min\left\{1, \left[(1+\frac{8\gamma_j\underline\psi_j^2}{\pi^2})/(1+\frac{8\gamma_j^*\underline\psi_j^{*^2}}{\pi^2})\right]^{0.75}\right\}$,
 where $E_j$ is calculated at $d_{\psi_j}\equiv 0$.
 The sampling schemes for $g$ and $H=h^2$ in steps PX1 and PX2 need to be updated accordingly.

The MDA algorithm for MMRM-st is an extension of Tang's algorithm  \cite{2017:tanga} for MMRM-n. It
can be easily adapted for MMRM-sn and MMRM-t. The details  are given in Appendices  \ref{adaptsn}  and \ref{adaptt}. 
In  MMRM-n and MMRM-t,  we can use the inverse Wishart distribution, Jeffrey's prior or the hierarchical prior
of  Huang and Wand \cite{huang:2013} for $\Sigma$. The latter two priors are noninformative and lead to  similar estimates. 
Whether the inverse Wishart distribution is informative or not depends on the choice of the prior parameters \cite{daniels:1999}.

\section{MDA algorithm for a more general MMRM}\label{mdapattern}
In model \eqref{mixed},  the fixed effects can vary freely over time, and there is a covariate by visit interaction for each covariate. It is a special case of the 
following  more general model 
\begin{equation}\label{mmrm_2}
 y_{ij}= \sum_{k=1}^{\Rs} \eta_{k} z_{ikj} + \sum_{k=1}^{q} \alpha_{kj} x_{ik}  + \xi_{ij} \text{ for } j=1,\ldots,p.
\end{equation}
The covariates can be split into two disjointed sets.
The  set $\mathcal{X}$ includes those  time invariant covariates $x_{ik}$'s whose effects vary over time. In the  set $\mathcal{Z}$, the value
of the  covariate  $z_{ikj}$ may change over time, but its effect  $\eta_k$ remains constant over time.  
As illustrated in Tang \cite{2017:tanga}, a covariate in   $\mathcal{X}$ can be transformed into $p$ covariates in $\mathcal{Z}$.
Either the set $\mathcal{X}$ or $\mathcal{Z}$ could be empty.
Whenever possible, we shall keep the covariates in $\mathcal{X}$  to improve the efficiency of the MDA algorithm \cite{2017:tanga}.

Similarly, model  \eqref{mmrm_2} can be factorized as  the following sequential regression models
\begin{eqnarray}\label{factor2} 
y_{ij} \sim N\left(\sum_{k=1}^{\Rs} \eta_{k} \underline{z}_{ikj} + \sum_{k=1}^{\Qs} \underline\alpha_{kj} x_{ik}+ \sum_{t=1}^{j-1}\beta_{jt}y_{it}, (d_i\gamma_j)^{-1}\right)
\text{ for } j=1,\ldots,p,
\end{eqnarray}
 where   $\underline{z}_{ikj}=z_{ikj}- \sum_{t=1}^{j-1} \beta_{jt} z_{ikt}$, and $\underline{\alpha}_{kj} =  \alpha_{kj} - \sum_{t=1}^{j-1} \beta_{jt} \alpha_{kt}$.

The MDA algorithm for model \eqref{mixed} can be adapted for model \eqref{mmrm_2} with the following minor modifications 
\begin{itemize}
\item[1.] In step P1, update  $\tilde{\xv}_{ij}$ as $(x_{i1},\ldots, x_{i\Qs}, \tilde{y}_{i1}, \ldots, \tilde{y}_{ij})'$, where $\tilde{y}_{ij}=y_{ij} -\sum_{k=1}^\Rss \eta_{k} z_{ijk}$.
As mentioned in Appendices  \ref{postwdint} and \ref{prepostnu}, some quantities [i.e. $y_{ij}^*$  and  $r_j$]
 shall  be defined according to model \eqref{mmrm_2} instead of model  \eqref{mixed}.
\item[2.] A step P1b is added after step P1 to draw $\etav$ by Gibbs sampler.  Let $\underline{\zv}_{j}$ be a $n_j\times \Rss$ matrix whose $ik$-th entry is  $\underline{z}_{ijk}$,
  $\underline{y}_{ij}=y_{ij} - \sum_{k=1}^{\Qs} \underline\alpha_{kj} x_{ik}- \sum_{t=1}^{j-1}\beta_{jt}y_{it}$,   $\underline{\yv}_{j}=(\underline{y}_{1j},\ldots, \underline{y}_{n_jj})'$,
 and $D_{ij}=\text{Diag}(d_1,\ldots,d_{n_j})$.
Under the prior $\etav=(\eta_1,\ldots,\eta_{\Rs})' \sim N(\etav_0, V_{\eta_0})$ [note that $\pi(\etav)\propto 1$ as $V_{\eta_0}^{-1}\rightarrow \zerov$],
the  posterior distribution of $\etav$ is normal
\begin{eqnarray*}
\begin{aligned}
\pi(\etav|\phiv,\nu,\wc_i\text{'s},d_i\text{'s},Y_o,Y_{m})  & \propto \pi(\etav) \exp\left[-\frac{1}{2}\sum_{j=1}^p \gamma_j(\underline{\yv}_j- \underline{\zv}_j\etav)'D_{ij}(\underline{\yv}_j- \underline{\zv}_j\etav)\right]  \\
& \propto  N(\hat\etav,\hat{V}_\eta),
\end{aligned}
\end{eqnarray*}
where $\hat{V}_{\eta} = [\sum_{j=1}^p \gamma_j\underline{\zv}_{j}'D_{ij}\underline{\zv}_{j}+V_{\eta_0}^{-1} ]^{-1}$,
 and $\hat\etav = \hat{V}_{\eta}[\sum_{j=1}^p \gamma_j \underline{\zv}_{j}' D_{ij}\underline{\yv}_j+V_{\eta_0}^{-1}\etav_0  ]$.
\end{itemize}

\section{Imputations of missing data due to dropout}\label{impdrop}
This section discusses the imputation of the dropout missing data in MMRM and  PMMs.
A common feature of these models is that  they assume the same marginal distribution for the outcomes prior to dropout.  
That is, the observed data ($\yv_{io}$'s) distributions are identical in PMMs and MMRM, 
and the intermittent missing data ($\yv_{im}$'s) are MAR. 

 We focus on  model \eqref{factor2} since model \eqref{factor} is a special case with $\eta_1=\ldots=\eta_\Rs=0$. 
The distribution of $\yv_{i}$ for subjects in pattern $s$ can be decomposed as 
\begin{equation}\label{dispat}
f(\yv_i) = \left[ \prod_{j=1}^s f(y_{ij}|z_{ij},\phiv,\etav,\nu)\right] H(\yv_{iw}|\yv_{io},\yv_{im},\phiv,\etav,\nu) 
\end{equation}
where  $H(\yv_{iw}|\yv_{io},\yv_{im},\phiv,\etav,\nu)$ is  the conditional distribution of $\yv_{iw}$ given $(\yv_{io},\yv_{im})$.
The missing data distribution $H(\cdot|\cdot)$  may depend on some additional parameters $\phiv_2$ to capture the deviation from MAR.
 Since  the observed data do not provide information about $\phiv_2$, we set $\phiv_2$ at some prespecified values, and suppress $\phiv_2$ in the notation \cite{2017:tanga}.

The joint likelihood for  $(s_i,\yv_{io},\yv_{im},\yv_{iw})$'s  can be written as the product 
of the  likelihood for the pattern $s_i$'s and the likelihood for  $( \yv_{io},\yv_{im},\yv_{iw})$'s. 
If the parameters indexing the two likelihoods are separable with independent priors, they are independent in the posterior distribution.
Therefore, the posterior distribution of  $(\yv_{im}\text{'s}, \yv_{iw}\text{'s}, \phiv, \etav,\nu)$  is given by 
\begin{eqnarray}\label{jointpos2}
\begin{aligned}
& \pi(\yv_{im}\text{'s}, \yv_{iw}\text{'s}, \phiv,\etav,\nu |Y_o, s_i\text{'s}) \\
& \,\, \propto 
\left[\pi(\phiv)\pi(\etav)\pi(\nu) \prod_{i=1}^n f(\yv_{io},\yv_{im}| \phiv,\etav,\nu)\right] \prod_{i=1}^{n_{tot}}  H(\yv_{iw}|\yv_{io},\yv_{im},\phiv,\etav,\nu).
\end{aligned}
\end{eqnarray}
The marginal posterior distribution of $(\yv_{im}\text{'s}, \phiv,\etav,\nu)$ is  $\pi(\phiv)\pi(\etav)\pi(\nu) \prod_{i=1}^n f(\yv_{io},\yv_{im}| \phiv,\etav,\nu)$, and they can be sampled using the  MDA algorithm $A$
through the introduction of the latent variables $(d_i,\wc_i)$'s. We can then impute $\yv_{iw}$'s from $H(\cdot|\cdot)$ after the MDA algorithm converges.
All arguments are essentially identical to that in Tang \cite{2017:tanga, 2017:tangb}.

\subsection{MMRM (MAR)}
 Under MAR, the conditional distribution of  $\yv_{iw}$   given the historical outcomes $(\yv_{io},\yv_{im})$ is the same between dropouts and subjects who remain in the trial.
 By  Lemma \ref{condAversion},  for subjects in pattern $s$, $\yv_{iw}$ can  be imputed sequentially from 
$$ y_{ij}=\sum_{k=1}^\Rss \eta_{k} \underline{z}_{ikj}+  \sum_{k=1}^q \underline\alpha_{kj} x_{ik}+ \underline\psi_{j} \wc_i + \sum_{t=1}^{j-1}\beta_{jt}y_{it}+ \frac{1}{\sqrt{d_i}} \varepsilon_j\, \text{ for }\, j\geq s+1,$$
where $\eta_k$'s, $\underline\alpha_{kj}$'s, $\underline\psi_j$'s, $\beta_{jt}$'s, $\gamma_j$'s, and $(\wc_i,d_i,\yv_{im})$ are the posterior samples from the MDA algorithm, and 
 $\varepsilon_{j} \sim N(0,\gamma_j^{-1})$. We can also impute $\yv_{iw}=(y_{i,s+1}, \ldots, y_{ip})$ in matrix form as 
$$\yv_{iw}^{\text{\tiny MAR}} = U_{s_{22}}^{-1} 
 \begin{bmatrix} \sum_{k=1}^\Rss \eta_{k} \underline{z}_{i,k,s+1}+  \sum_{k=1}^q \underline\alpha_{k,s+1} x_{ik}+ \underline\psi_{s+1} \wc_i +  \sum_{t=1}^{s}\beta_{s+1,t}y_{it}+ \frac{\varepsilon_{s+1}}{\sqrt{d_i}}\\
\ldots \\
\sum_{k=1}^\Rss \eta_{k} \underline{z}_{ikp}+  \sum_{k=1}^q \underline\alpha_{kp} x_{ik}+ \underline\psi_{p} \wc_i + \sum_{t=1}^{s}\beta_{pt}y_{it}+ \frac{\varepsilon_p}{\sqrt{d_i}} \end{bmatrix},
   $$
where $U_{s_{22}}$ is defined in Lemma \ref{condAversion}.

\subsection{Delta-adjusted imputation}
In the delta-adjusted PMMs, the mean response at visit $j>s$ among subjects in treatment group $g$, pattern $s$
will be shifted by a fixed value $\Delta_{sgj}$   compared to those who remain on the treatment at visit $j$ \cite{2013:mallinckrodta, 2017:tangb}
$$ y_{ij}=\Delta_{sgj}+\sum_{k=1}^\Rss \eta_{k} \underline{z}_{ikj}+  \sum_{k=1}^q \underline\alpha_{kj} x_{ik}+ \underline\psi_{j} \wc_i + \sum_{t=1}^{j-1}\beta_{jt}y_{it}+ \frac{1}{\sqrt{d_i}} \varepsilon_j\, \text{ for }\,j\geq s+1.$$
The imputed values can be conveniently obtained from that under MAR as 
$$ \yv_{iw}^{\text{\tiny DEL}} =\yv_{iw}^{\text{\tiny MAR}}+ U_{s_{22}}^{-1}  (\Delta_{sg,s+1},\ldots,\Delta_{sgp})'.$$
Sensitivity analysis can be performed by varying the  parameters $\Delta_{sgt}$'s.  To reduce the amount of sensitivity parameters, we set $\Delta_{sgt}=\Delta_g$. 
But other choices are possible \cite{ 2017:tangb}. 
The delta adjustment is applied  by conditioning on the historical outcomes $(y_{i1},\ldots,y_{is})$.  The  adjustment 
 can also be performed  without conditioning on the historical outcomes \cite{2017:tangb}, i.e., $\yv_{iw}^{\text{\tiny DEL}} =\yv_{iw}^{\text{\tiny MAR}}+ (\Delta_{sg,s+1},\ldots,\Delta_{sgp})'$.

The delta-adjusted imputation forms the basis of  the {\it tipping point}  analysis. The tipping point analysis assesses how severe the deviation from MAR can be in order to overturn the MAR-based conclusion. It
 has been popularly used in new drug applications \cite{ich:2017}. 
 The delta adjustment can be applied only to the experimental arm \cite{2017:tanga, 2017:tangb} by assuming MAR in the control arm ($\Delta_0=0$)  or to both arms \cite{permutt:2016, tang:2019}.
The MI analysis is repeated over a  sequence of prespecified values for $\Delta_1$ given $\Delta_0=0$  or over a range of pre-specified values for $(\Delta_0, \Delta_1)$ in order to find the 
region in which the treatment comparison becomes statistically insignificant.
If the insignificance region  is deemed clinically implausible,  the primary conclusion is said to be robust to deviations from MAR.

\subsection{Control-based imputation}
The control-based imputation assumes  that after dropout, the future statistical behavior among subjects in the experimental arm is similar to that of control subjects,
and that the missing data are MAR in the control group.
Therefore, the missing data $y_{iw}$'s in the experimental arm can be imputed by borrowing information from the control arm.

The idea was firstly proposed in the seminal paper by Little and Yau \cite{little:1996}, and was later extended by a number of authors  \cite{2013:carpenter,2013:mallinckrodta, lu:2015d, 2017:tanga, 2017:tangb,2017:tangc,tang:2018, tang:2017e, tang:2019} for different types of response variables.
Several popular control-based imputation strategies include the jump to reference (J2R), copy increment in reference (CIR), and copy reference (CR).

\subsubsection{Jump to reference (J2R)}
The J2R approach assumes that once subjects in the experimental arm cease the treatment, their mean responses jump to that of the control subjects.
The model essentially assumes that immediately upon withdrawal from the experimental treatment, all benefit from the treatment is gone \cite{2013:carpenter, 2013:mallinckrodta}.

In J2R, $\yv_{iw}$ can be imputed as 
 $$\yv_{iw}^{\text{\tiny J2R}}  =\yv_{iw}^{\text{\tiny MAR}}- (\delta_{s+1},\ldots,\delta_p)'\,g_i.$$
Let's demonstrate the fact by using MMRM-st as an example.
Suppose the distribution of $\yv_i$ in MMRM is $ST[(\mu_{i1}^{\text{\tiny{C}}}+\delta_1 g_i,\ldots,\mu_{ip}^{\text{\tiny{C}}}+\delta_pg_i), \psiv, \Sigma,\nu]$, where 
$\mu_{ij}^{\text{\tiny{C}}}=\sum_{k=1}^r \eta_{k} z_{ijk} + \sum_{k=1}^{q-1} \alpha_{kj} x_{ik}$ 
and $\delta_j$ is the treatment effect at visit $j$.
In J2R, the distribution of $\yv_i$ is  $ST[(\mu_{i1}^{\text{\tiny{C}}}+\delta_1 g_i,\ldots,\mu_{is}^{\text{\tiny{C}}}+\delta_sg_i, \mu_{i,s+1}^{\text{\tiny{C}}},\ldots,\mu_{ip}^{\text{\tiny{C}}}), \psiv, \Sigma,\nu]$, 
and the treatment effect vector is $(\delta_1,\ldots,\delta_s,0,\ldots,0)$ in pattern $s$. 
By Lemma \ref{condAversion}c, the conditional distributions of $\yv_{iw}$ given $(y_{i1},\ldots,y_{is})$ in MMRM and J2R differ only in the location parameters, and the difference is $(\delta_{s+1}g_i,\ldots,\delta_pg_i)'$.

\subsubsection{ Copy Increment in Reference (CIR)}
In CIR, the mean profile after dropout in the experimental arm is assumed to be parallel to that of control subjects. The
  treatment benefit prior to withdrawal is maintained in  CIR. It is suitable for modeling the effectiveness of a disease modifying treatment \cite{2013:carpenter, 2013:mallinckrodta}. 
 In pattern $s$, the distribution of $\yv_i$ is  $ST[(\mu_{i1}^{\text{\tiny{C}}}+\delta_1 g_i,\ldots,\mu_{is}^{\text{\tiny{C}}}+\delta_s g_i, \mu_{i,s+1}^{\text{\tiny{C}}}+ \delta_s g_i,\ldots,
 \mu_{ip}^{\text{\tiny{C}}}+ \delta_s g_i), \psiv, \Sigma,\nu]$, and
 the treatment effect vector is $(\delta_1,\ldots,\delta_s,\delta_s,\ldots,\delta_s)$.  
We can impute $\yv_{iw}$ as
$$\yv_{iw}^{\text{\tiny CIR}} = \yv_{iw}^{\text{\tiny J2R}} + (\delta_s,\ldots,\delta_s)'\,g_i = \yv_{iw}^{\text{\tiny MAR}}- (\delta_{s+1}-\delta_s,\ldots,\delta_p-\delta_s)'\,g_i.$$

\subsubsection{Copy Reference (CR)}
Under the  CR assumption, the missing data distribution of $\yv_{iw}$ given $(y_{i1},\ldots,y_{is})$ among  dropouts from the experimental arm  
is the same as that of control subjects. 
The missing data distribution for dropouts from the experimental arm  can be obtained on basis of
Lemma \ref{condAversion} by assuming that they received the control treatment since randomization, and had
the response distribution  $ST[(\mu_{i1}^{\text{\tiny{C}}},\ldots,\mu_{ip}^{\text{\tiny{C}}})', \Sigma, \psi,\nu]$. The true joint distribution of $\yv_i=(y_{i1},\ldots,y_{ip})'$ is complicated.
For the purpose of missing data imputations, we can firstly draw $(d_i^*,\wc_i^*)$ given $(y_{i1},\ldots,y_{is})$ using Lemma \ref{condAversion}a, and then 
impute $\yv_{iw}$ given  $(y_{i1},\ldots,y_{is},d_i^*,\wc_i^*)$ using Lemma \ref{condAversion}b or \ref{condAversion}c.

In CR,  $(d_i^*,\wc_i^*)$ among dropouts from the experimental arm needs to be drawn on basis of the control mean. 
But in MMRM (MAR), delta-adjusted imputation, J2R and CIR, there is no need to regenerate  $(d_i^*,\wc_i^*)$ since it has the same posterior distribution as $(d_i,\wc_i)$ 
from the MDA algorithm.

\section{Numerical examples}\label{numexam}

\begin{figure}[htb]
\subfigure[data with $30$ subjects simulated from $y_{i} \sim SN(0.5 g_i+0.5x_{i},0.25,0.5)$]
{
\includegraphics[scale=0.35]{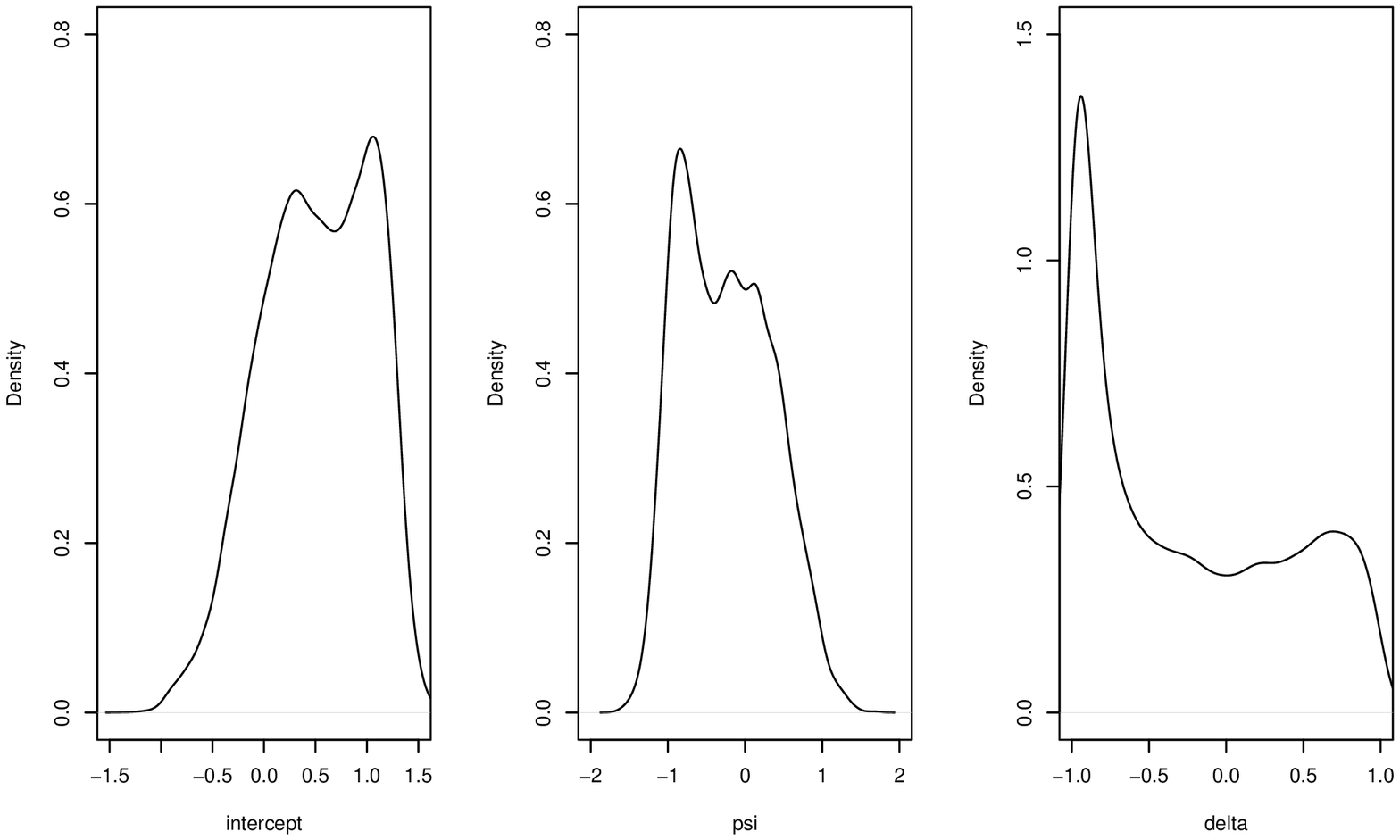}\label{mulmod1}
}\,\,\,\,\,
\subfigure[data with $30$ subjects simulated from $y_{i} \sim SN (0.5 g_i+0.5x_{i},1, 0.5)$]
{
\includegraphics[scale=0.35]{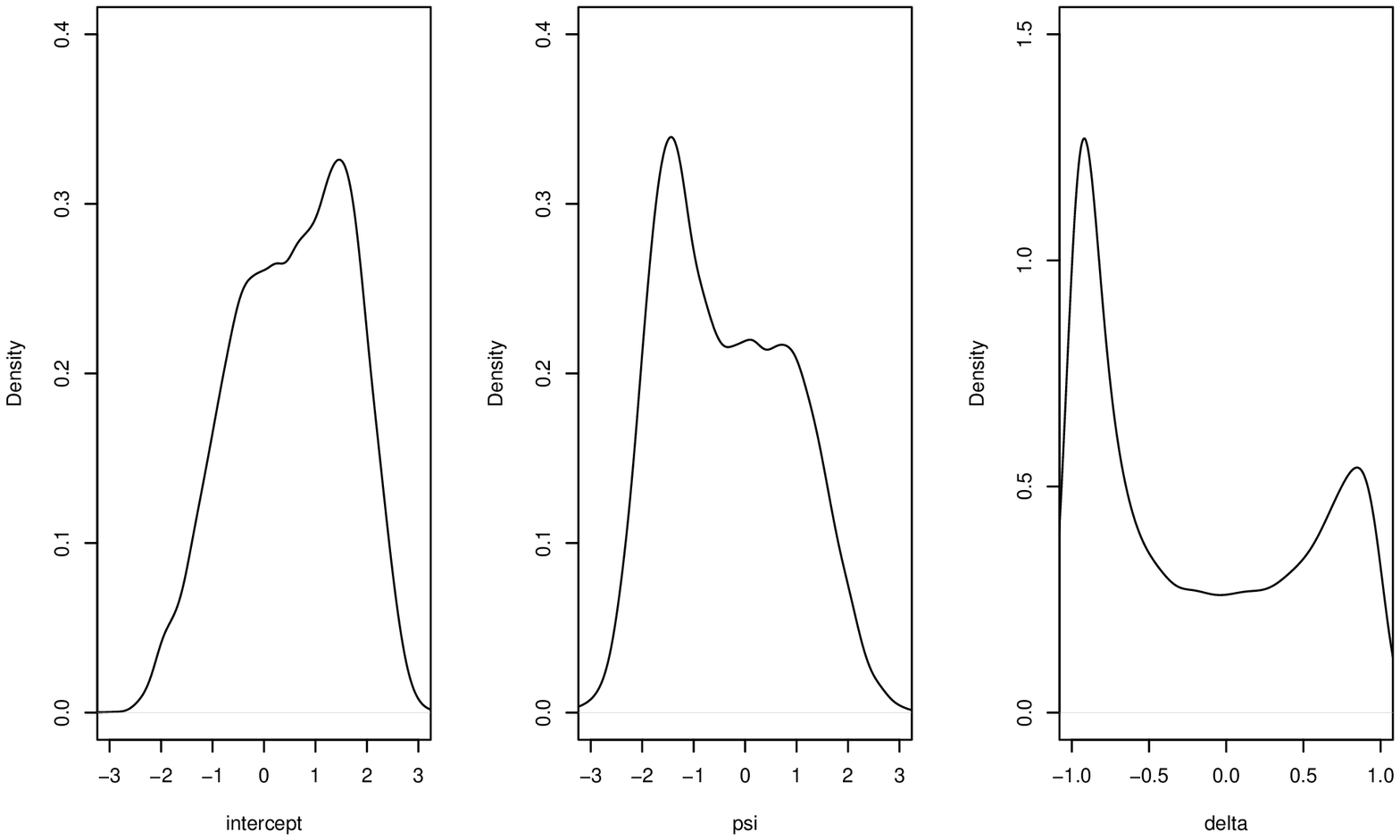}\label{mulmod2}
}
\caption{Posterior densities of the intercept, $\psi$ and $\delta=\frac{\lambda}{\sqrt{1+\lambda^2}}$ in the univariate SN regression}\label{multmodalitysn}
\end{figure}

\subsection{Multimodality in the SN regression }
In the univariate SN regression, the expected Fisher information matrix  is singular\cite{AZZALINI:1985} at $\lambda=\psi/\sigma=0$. 
As a consequence, the empirical distribution of the MLE,
and the posterior  distribution of the parameter  in Bayesian  inference are often bimodal  \cite{pewsey:2000, Arellano:2008} when $\lambda$ is near $0$. 
 Liseo and Parisi \cite{liseo:2013} raises a concern that the Gibbs sampler chain can easily get stuck in one of the modes for multimodal posterior distributions.

This does not appear to be a concern in our algorithm possibly because we sample $(\thetav_j,\gamma_j)$'s simultaneously using the blocking scheme.  The Gibbs sampler can be highly inefficient if the intercept and $\underline\psi_j$  are
sampled separately because they are highly correlated. For illustrative purposes, we generate two datasets of size $n=30$  from $y_i\sim SN(\alpha_1+\alpha_2 x_i+\alpha_3 g_i, \sigma^2,\psi)$.
Figure \ref{multmodalitysn} plots the posterior densities of $\alpha_1$, $\psi$ and $\delta=\frac{\lambda}{\sqrt{1+\lambda^2}}$. They are clearly bimodal or multimodal for both datasets.

\subsection{Analysis of an antidepressant trial using controlled  imputation}\label{antidep}
The antidepressant trial has been analyzed by several authors to illustrate the missing data methodologies   \cite{2013:mallinckrodta, 2015a:tang, 2017:tanga, 2017:tangb, tang:2019,2017:tangc}.
The Hamilton 17-item rating scale for depression  ($\text{HAMD}_{17}$)  is collected at baseline and weeks 1, 2, 4, 6. The
dataset consists of 84 subjects on the experimental treatment and 88 subjects on placebo. The number of subjects who discontinue the trial early is  $20$ ($24\%$)  in the experimental arm, and $23$ ($26\%$) in the placebo arm. 

The purpose of the analysis is to estimate the effect of the experimental product   compared to placebo on the improvement  in $\text{HAMD}_{17}$ from baseline to week $6$. 
We impute the missing response under MAR and MNAR by MMRM-n, MMRM-t, MMRM-sn and MMRM-st. 
The covariate set $\mathcal{X}$ includes the intercept, baseline $\text{HAMD}_{17}$ score $y_{i0}$ and treatment status $g_i$. The covariate  set $\mathcal{Z}$ is empty. 
In each model,  $m=10,000$ datasets are imputed from every $100$th iteration after a burn-in period of $100,000$ iterations.
The trace plots and autocorrelation function (ACF)  plots indicate approximate convergence of these MDA algorithms. 
In practice, it is prudent to use a long burn-in period to ensure that the Markov chain reaches the stationary distribution, and this is particularly important 
in the pharmaceutical industry where the  analysis is done by programmers without much knowledge about the Bayesian analysis \cite{2017:tanga}.
 We analyze the outcome  at week $6$  by the analysis of covariance (ANCOVA)  for each imputed dataset, and the results are combined for inference using Rubin's rule \cite{barnard:1999}. 

We employ the deviance information criterion (DIC  \cite{spiegelhalter:2002}) to compare the four MMRMs. The DIC is defined as
\begin{eqnarray*}
\begin{aligned}
 \text{DIC} & =D(\hat\etav,\hat\phiv,\hat\nu)+2\text{pD}=D(\hat\etav,\hat\phiv,\hat\nu)+ 2[\bar{D}(\etav,  \phiv,\nu)-D(\hat\etav,\hat\phiv,\hat\nu)] \\
& = 2m^{-1}\sum_{b=1}^m D(\etav^{(b)},  \phiv^{(b)},\nu^{(b)})- D(\hat\etav,\hat\phiv,\hat\nu),
\end{aligned}
\end{eqnarray*}
where $D(\etav,  \phiv,\nu) = -2\sum_{i=1}^n \log[ f(\yv_{io}| \etav,  \phiv,\nu)]$,   
 $(\etav^{(b)},  \phiv^{(b)},\nu^{(b)})$ is the $b$-th posterior sample collected after the burn-in period, 
$(\hat\etav,\hat\phiv,\hat\nu)=m^{-1}\sum_{b=1}^m (\etav^{(b)} ,\phiv^{(b)},\nu^{(b)})$, and 
$\bar{D}(\etav,  \phiv,\nu)= m^{-1}\sum_{b=1}^m D\left(\etav^{(b)},  \phiv^{(b)},\nu^{(b)} \right)$. 
In DIC, $D(\hat\etav,\hat\phiv,\hat\nu)$ measures the model fit while  $\text{pD}$ estimates
 the effective number of parameters or model complexity  \cite{spiegelhalter:2002}. Overall, a smaller DIC indicates a better model fit. 

  \begin{table}[htb]
\begin{center}
\tiny
\begin{tabular}{l@{\extracolsep{5pt}}c@{\extracolsep{5pt}}c@{\extracolsep{5pt}}c@{\extracolsep{5pt}}c@{\extracolsep{5pt}}c@{\extracolsep{5pt}}c@{\extracolsep{5pt}}c@{\extracolsep{5pt}}c@{\extracolsep{5pt}}c@{\extracolsep{5pt}}c@{\extracolsep{5pt}}c@{\extracolsep{5pt}}c@{\extracolsep{5pt}}ccccc} \\\hline  
& \multicolumn{2}{c}{MMRM-n} & \multicolumn{2}{c}{MMRM-t}  & \multicolumn{2}{c}{MMRM-sn} & \multicolumn{2}{c}{MMRM-st} \\\cline{2-3}\cline{4-5}\cline{6-7}\cline{8-9}
 & mean$\pm$ SE & t (pvalue) &     mean$\pm$ SE & t (pvalue)  & mean$\pm$ SE & t (pvalue)  & mean$\pm$ SE & t (pvalue)  \\\hline 
    MAR&$   -2.80\pm 1.11$&$ -2.54\,(0.012)$&$   -2.81\pm 1.13$&$ -2.49\,(0.014)$&$   -2.80\pm 1.14$&$ -2.47\,(0.015)$&$   -2.81\pm 1.11$&$ -2.54\,(0.012)$\\
                                                   J2R&$   -2.13\pm 1.12$&$ -1.90\,(0.059)$&$   -2.11\pm 1.14$&$ -1.85\,(0.066)$&$   -2.14\pm 1.15$&$ -1.87\,(0.064)$&$   -2.16\pm 1.12$&$ -1.93\,(0.056)$\\
                                                   CR&$   -2.37\pm 1.10$&$ -2.15\,(0.033)$&$   -2.37\pm 1.11$&$ -2.13\,(0.035)$&$   -2.37\pm 1.12$&$ -2.11\,(0.036)$&$   -2.38\pm 1.10$&$ -2.16\,(0.032)$\\ 
                                                   CIR&$   -2.45\pm 1.10$&$ -2.23\,(0.027)$&$   -2.46\pm 1.12$&$ -2.20\,(0.030)$&$   -2.45\pm 1.13$&$ -2.17\,(0.032)$&$   -2.47\pm 1.10$&$ -2.25\,(0.026)$\\
                                                   DEL$^{(a)}$&$   -2.05\pm 1.13$&$ -1.82\,(0.071)$&$   -2.05\pm 1.15$&$ -1.78\,(0.076)$&$   -2.06\pm 1.16$&$ -1.78\,(0.078)$&$   -2.07\pm 1.13$&$ -1.83\,(0.069)$\\
\hline
 \end{tabular} \caption{MI treatment effect estimates at week $6$ in sensitivity analysis of an antidepressant trial using controlled pattern imputations: \newline
$^{(a)}$ a delta adjustment of $-2$ is applied to subjects in the experimental arm after treatment discontinuation. MAR is assumed in the placebo arm.
 }\label{depres}
\end{center}
\end{table}

 Table \ref{depres} reports the MI results. The DIC is $3526.97$ for MMRM-n, $3528.58$ for MMRM-t, $3514.55$ for MMRM-sn and $3514.43$ for MMRM-st.
 MMRM-st appears to fit the data slightly better and give slightly more significant treatment effect estimates
than MMRM-n, MMRM-t and MMRM-sn.

\begin{figure}[htb]
\centering
\subfigure[pvalue$<0.05$ above the dashed horizontal line]
{
\includegraphics[scale=0.525]{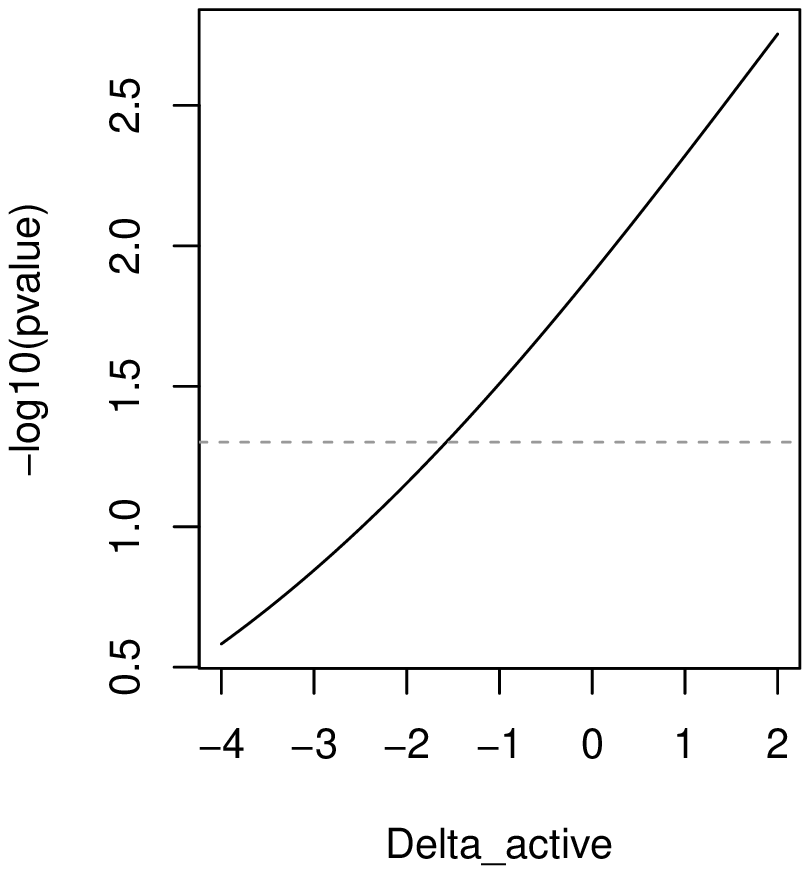}\label{figtip1}
}
\subfigure[The symbols indicate the range of pvalue: `v' pvalue$<0.0001$, `x' pvalue$<0.001$, `o' pvalue $<0.01$, `*' pvalue$<0.05$]
{
\includegraphics[scale=0.525]{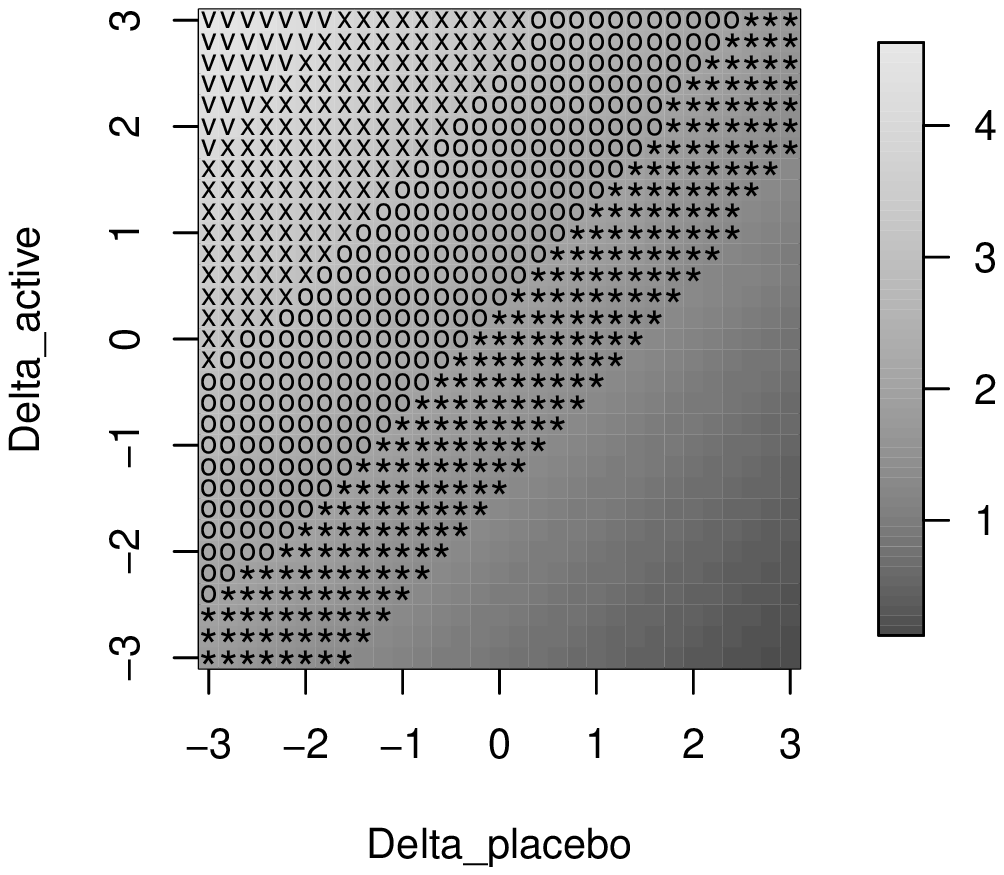}\label{figtip2}
}
\caption{Plot of -$\log_{10}(\text{pvalue})$ in the tipping point analysis of an  antidepressant trial with delta adjustment in the experimental arm (left) and in both treatment groups (right) }\label{figtip}
\end{figure}

As will be discussed in the last section, we suggest conducting  the missing value imputation based directly on MMRM-st without a model selection in large trials.
Below we illustrate the tipping point analysis on basis of MMRM-st.
Figure \ref{figtip1} plots the result when the  adjustment is applied only in the experimental arm (i.e. MAR   in the placebo arm).
The treatment comparison becomes insignificant (pvalue$>0.05$) if the mean response among dropouts from the experimental arm is at least $1.4$ point worse at each visit compared to subjects who remain on the experimental treatment.
Figure \ref{figtip2} plots the analysis with the delta adjustment in both arms. The treatment effect becomes insignificant only in a small region where $\Delta_0-\Delta_1\geq 1.4$ roughly holds.

\subsection{ Framingham cholesterol data}\label{cholana}
We  analyze the Framingham cholesterol data to assess the robustness of MMRM-n, MMRM-t, MMRM-sn and MMRM-st in the presence of outliers. 
The data were first explored by Zhang and Davidian \cite{zhang:2001} to characterize changes in the cholesterol level over time, and assess the effect of age  and gender.
Two hundred  subjects  are randomly selected from the Framingham study. 
The cholesterol levels are measured at the beginning of the study and then every $2$ years for $10$ years. 

In the literature\cite{zhang:2001,ghidey:2004,jara:2008,lachos:2009,cabral:2012}, this dataset was typically fitted by a linear growth (LG) model with  baseline age and gender as fixed effects and subject-specific random intercept and slopes,
where the random effects and/or random errors are  modeled by non-normal distributions. 
We provide an alternative approach to analyze the data
$$ y_{ij}= \eta_1 + \eta_2 \,t_{ij} +\eta_3 \,\text{sex}_i+ \eta_4 \,\text{age}_i +\xi_{ij},$$
where $y_{ij}$ is the cholesterol level divided by $100$ at visit $j$ for
subject $i$,  and $t_{ij}$ is $(\text{time}-5)/10$ with time measured in years from baseline.
To compare the  fixed effect estimates with those reported in the literature, we put $4$ covariates $\{1, t_{ij}, \text{sex}_i, \text{age}_i\}$ in the set $\mathcal{Z}$, and no covariate in  $\mathcal{X}$.
Another  approach is to set $\mathcal{X}=\{1, \text{sex}_i, \text{age}_i\}$  and $\mathcal{Z}= \{ t_{ij}\}$, and it makes fewer assumptions on the relationship of age and gender with the cholesterol level.
The within subject dependence is modeled by the multivariate normal, t, SN or ST distributions. 
Our model is more general than the LG  model in that we don't  assume a structured  covariance matrix. 

\begin{figure}[htb]
\centering
\includegraphics[scale=0.6]{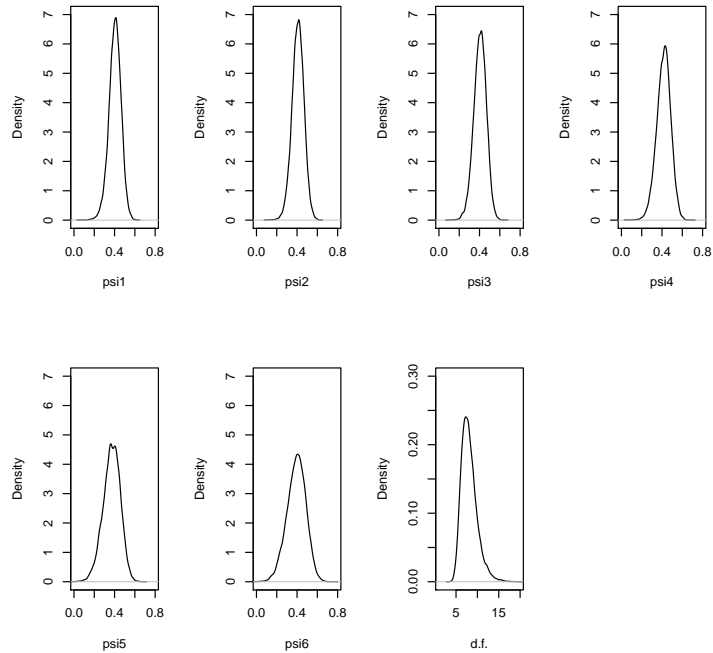}
\caption{Marginal posterior densities for the skewness ($\psi_1,\ldots,\psi_6$) and df ($\nu$)  parameters in the analysis of the cholesterol raw data using MMRM-st}\label{histchol}
\end{figure}

In the MDA algorithm, we collect  $20,000$ posterior samples from every $100$th iteration after a burn-in period of 
$100,000$ iterations. The  convergence of the Markov chain is evidenced by the
trace plots and ACF plots  in all four models.

Table \ref{cholres} displays the parameter estimates  and DIC. 
 According to the DIC criterion, MMRM-st provides the best fit to the raw data. 
Figure \ref{histchol} plots the marginal posterior densities for the skewness and df parameters in MMRM-st.
The posterior samples of $\nu$ concentrate in the interval $[5,15]$, indicating heavy tails in the observed data.
The posterior densities for $\psi_1,\ldots,\psi_6$ all concentrate in the interval $[0.1,0.7]$, indicating the skewness in the cholesterol level  at each visit.
As shown in Table  \ref{cholres}, the $95\%$ credible intervals for $\underline\psi_3,\ldots,\underline\psi_6$ cover $0$, evidencing that the skewness of the  cholesterol level
reduces after adjusting for the historical outcomes at  previous visits. 

In these MMRMs, the regression coefficients (particularly the intercept $\eta_1$)  do not have the same interpretation because the latent variable $\wc\sim N^+(0,1)$ does not have zero mean. 
Let $E_i=\eta_1 + \eta_2 \,t_{ij} +\eta_3 \,\text{sex}_i+ \eta_4 \,\text{age}_i$. 
In both MMRM-n and MMRM-t, the mean response is assumed to be  constant over time,  $\Ev_{i_\text{n}}=\Ev_{i_\text{t}} =(E_i,E_i,E_i,E_i, E_i,E_i)'$.
In MMRM-sn and MMRM-st, the mean response is not constrained to be the same  across visits. 
 The mean response profile is given by 
$\Ev_{i_\text{sn}}=\Ev_{i_\text{n}} + \sqrt{\frac{2}{\pi}}(\psi_1,\ldots,\psi_6)'$ in MMRM-sn, and
 $\Ev_{i_\text{st}}=\Ev_{i_\text{n}} + \sqrt{\frac{\nu}{\pi}} \frac{\Gamma(\frac{\nu-1}{2})}{ \Gamma(\frac{\nu}{2})}(\psi_1,\ldots,\psi_6)'$ in MMRM-st.
The Bayes estimates of the fixed effects in MMRM-n and MMRM-t are close to the MLE from the normal LG model  reported by Zhang and Davidian \cite{zhang:2001},
and MMRM-t gives slightly narrower credible intervals for the fixed effects than MMRM-n.
Lachos {\it et al} \cite{lachos:2009} analyzes the data using the robust LG  model with SN (or ST) random intercept and slope, and normal (or t) random error,
which can be roughly viewed as the submodels of the robust MMRM \eqref{mmrm_2} with certain constraints on the covariance parameters  $\Sigma$ and the skewness parameters
$\psiv=(\psi_1,\ldots,\psi_p)'$.
The estimated age and gender effects [$\hat\eta_3=-0.057$ (SD: $0.049$), $\hat\eta_4=0.014$ (SD: $0.003$)] from the LG model with SN random effects are
 close to that from MMSM-sn, while the LG model with ST random effects  gives similar estimates of the age and gender effects  [$\hat\eta_3=-0.062$ (SD: $0.045$), $\hat\eta_4=0.014$ (SD: $0.003$)] to MMRM-st.
The estimate and interpretation of the intercept $\eta_1$ and slope $\eta_2$ parameters are different in these models since the skewed random variables do not have zero mean.
Lachos {\it et al} \cite{lachos:2010a} assesses the performance of the robust LG models in predicting future responses. 
The robust MMRMs show comparable  prediction performance, and the results are not shown due to limited space.

We evaluate  the robustness of these MMRMs  through the influence of the outliers on the parameter estimates.  
For simplicity, the outlier values are generated by replacing $y_{ij}$ with $y_{ij}=y_{ij}+8$ for $j=1,\ldots,6$ in the first two subjects.  
The result is also displayed in  Table \ref{cholres}. In MMRM-n, the estimates of the regression coefficients and their $95\%$ credible intervals
for the between-subject covariates (i.e. intercept $\eta_1$, sex $\eta_3$, age $\eta_4$) change noticeably after the introduction of the outliers,
but the estimate of the time effect $\eta_2$  is little changed.
Similar behavior is observed in the maximum likelihood inference by Zhang and Davidian \cite{zhang:2001}.
In MMRM-sn, the outliers influence the estimates of the skewness parameters $\hat{\underline\psi_j}$'s and the $95\%$ credit intervals for the regression coefficients.
MMRM-t provides robust parameter estimates except that the estimated df parameter gets smaller,  indicating heavier tails in the presence of outliers.
 In MMRM-st,  the outliers affect the estimates of the df and skewness parameters, but the estimation of the covariate effects is much less sensitive to the outliers.

\begin{landscape}
 \begin{table}[htb]
\footnotesize
\begin{center}
\begin{tabular}{lccccccccccccccccc} \\\hline  
 & MMRM-n & MMRM-t &  MMRM-sn & MMRM-st   \\\hline 
\multicolumn{5}{c}{Framingham cholesterol raw data}\\
                                     $\eta_1$&$   1.647\pm0.148\,\, [   1.357,   1.937]$&$   1.567\pm0.141\,\, [   1.295,   1.846]$&$   1.395\pm0.140\,\, [   1.121,   1.668]$&$   1.414\pm0.133\,\, [   1.148,   1.674]$\\
                                     $\eta_2$&$   0.275\pm0.025\,\, [   0.227,   0.324]$&$   0.277\pm0.024\,\, [   0.230,   0.324]$&$   0.332\pm0.091\,\, [   0.146,   0.500]$&$   0.296\pm0.082\,\, [   0.128,   0.450]$\\
                                     $\eta_3$&$  -0.063\pm0.054\,\, [  -0.168,   0.043]$&$  -0.067\pm0.051\,\, [  -0.167,   0.033]$&$  -0.056\pm0.049\,\, [  -0.152,   0.039]$&$  -0.063\pm0.046\,\, [  -0.154,   0.026]$\\
                                     $\eta_4$&$   0.017\pm0.003\,\, [   0.011,   0.024]$&$   0.018\pm0.003\,\, [   0.012,   0.025]$&$   0.014\pm0.003\,\, [   0.008,   0.021]$&$   0.014\pm0.003\,\, [   0.008,   0.020]$\\
                                     $\underline\psi_1$& & &$   0.494\pm0.065\,\, [   0.358,   0.609]$&$   0.404\pm0.061\,\, [   0.277,   0.517]$\\                                                                        
                                     $\underline\psi_2$& & &$   0.334\pm0.123\,\, [   0.097,   0.584]$&$   0.271\pm0.091\,\, [   0.105,   0.460]$\\                                                                        
                                     $\underline\psi_3$& & &$   0.051\pm0.120\,\, [  -0.172,   0.292]$&$   0.073\pm0.098\,\, [  -0.115,   0.269]$\\                                                                        
                                     $\underline\psi_4$& & &$   0.116\pm0.145\,\, [  -0.148,   0.407]$&$   0.109\pm0.105\,\, [  -0.087,   0.320]$\\                                                                        
                                     $\underline\psi_5$& & &$  -0.225\pm0.143\,\, [  -0.478,   0.083]$&$  -0.136\pm0.116\,\, [  -0.353,   0.097]$\\                                                                        
                                     $\underline\psi_6$& & &$  -0.025\pm0.168\,\, [  -0.348,   0.311]$&$   0.023\pm0.132\,\, [  -0.234,   0.288]$\\                                                                        
                                     $\nu$& &$   8.532\pm2.058\,\, [   5.472,  13.458]$& &$   8.091\pm1.869\,\, [   5.275,  12.542]$\\                                                                                     
DIC  & $348.65$ & $310.56$ & $339.285$ & $296.01$ \\     
\\                                      

\multicolumn{5}{c}{Framingham  cholesterol data with outliers generated  in the first two subjects}\\
                                   $\eta_1$&$   1.730\pm0.356\,\, [   1.034,   2.425]$&$   1.545\pm0.147\,\, [   1.258,   1.834]$&$   1.397\pm0.190\,\, [   1.011,   1.756]$&$   1.420\pm0.139\,\, [   1.144,   1.689]$\\
                                     $\eta_2$&$   0.276\pm0.024\,\, [   0.229,   0.324]$&$   0.279\pm0.024\,\, [   0.234,   0.326]$&$   0.322\pm0.032\,\, [   0.260,   0.385]$&$   0.292\pm0.056\,\, [   0.181,   0.399]$\\
                                     $\eta_3$&$  -0.019\pm0.130\,\, [  -0.273,   0.236]$&$  -0.067\pm0.053\,\, [  -0.171,   0.037]$&$  -0.050\pm0.070\,\, [  -0.189,   0.089]$&$  -0.065\pm0.047\,\, [  -0.157,   0.028]$\\
                                     $\eta_4$&$   0.017\pm0.008\,\, [   0.001,   0.033]$&$   0.018\pm0.004\,\, [   0.012,   0.025]$&$   0.009\pm0.004\,\, [   0.000,   0.017]$&$   0.013\pm0.003\,\, [   0.006,   0.019]$\\
                                     $\underline\psi_1$& & &$   1.159\pm0.067\,\, [   1.035,   1.296]$&$   0.468\pm0.056\,\, [   0.358,   0.578]$\\                                                                        
                                     $\underline\psi_2$& & &$   1.137\pm0.249\,\, [   0.592,   1.613]$&$   0.350\pm0.101\,\, [   0.167,   0.563]$\\                                                                        
                                     $\underline\psi_3$& & &$   0.197\pm0.390\,\, [  -0.522,   1.031]$&$   0.107\pm0.113\,\, [  -0.111,   0.341]$\\                                                                        
                                     $\underline\psi_4$& & &$   0.321\pm0.461\,\, [  -0.484,   1.307]$&$   0.138\pm0.115\,\, [  -0.074,   0.379]$\\                                                                        
                                     $\underline\psi_5$& & &$  -0.060\pm0.447\,\, [  -0.962,   0.903]$&$  -0.131\pm0.128\,\, [  -0.368,   0.132]$\\                                                                        
                                     $\underline\psi_6$& & &$  -0.059\pm0.424\,\, [  -0.966,   0.801]$&$   0.020\pm0.131\,\, [  -0.240,   0.283]$\\                                                                        
                                     $\nu$& &$   4.957\pm0.756\,\, [   3.648,   6.596]$& &$   5.318\pm0.853\,\, [   3.873,   7.225]$\\                                                                                     

DIC                          &     $699.60$  & $402.50$ & $544.95$ & $371.94$ \\    
\hline
 \end{tabular} \caption{DIC and posterior mean $\pm$ standard deviation [$95\%$ credible interval] for the model parameters in the analysis of the Framingham cholesterol raw data and data  with outliers generated 
 in the first two subjects :\newline
[1] the estimates of the variance parameters (i.e. $\beta_{jk}$'s and $\gamma_j$'s) are omitted due to the limited space.
 }\label{cholres}
\end{center}
\end{table}
\end{landscape}

 \section{Discussion}\label{discussion}
We consider robust inference for skewed and/or heavy-tailed  longitudinal data  using MMRM-st, MMRM-sn or MMRM-t. These robust regressions have some undesirable attributes,
and the posterior distributions can be improper with infinite estimates  for some model parameters under diffuse priors. 
We use the PC prior  for  the df parameter, Huang-Wand's \cite{huang:2013} hierarchical prior for $\Sigma$, and reference prior for the individual skewness parameter $\underline\psi_j$ of $\underline{y}_{ij}=y_{ij}-\sum_{t=1}^{j-1}\beta_{jt}y_{it}$.   
 An efficient MDA algorithm 
is developed for Bayesian inference and missing data imputation.
In practice,  one may specify a different prior that reflects the existing knowledge or has better statistical properties.  The MDA algorithm can be modified accordingly.
 For example, if a non-conjugate prior is used for the skewness parameters, they can be drawn via an independent or random walk Metropolis sampler\cite{chib:1995} with  candidates generated by the proposed Gibbs scheme.

In clinical trials, usually only a few important covariates (e.g. baseline response, stratification factors) are  included in the model \cite{chmp:2015,senn:1994}. These covariates are 
typically completely observed. In case there are some missing covariates, the MDA algorithm can be adapted to impute both the missing covariates and responses based on their joint distribution \cite{tang:2019}.  
In Lu \cite{lu:2010dd}, the baseline covariates are constrained to have the same mean across treatment groups in randomized trials.
Relaxing this constraint simplifies the algorithm without incurring efficiency loss in randomized trials (simulations unreported here), and  makes it also suitable for studies with  baseline  imbalance.

The MDA algorithm is used to perform the controlled  imputations for MNAR sensitivity analyses of longitudinal clinical trials. 
 The assumptions about missing data  are  untestable given only the observed data \cite{NRC:2010}.
A control-based strategy (CR, J2R or CIR) can be selected  according to the drug mechanism of action (i.e. will the treatment benefit disappear after treatment discontinuation? how long will it take for  the benefit to disappear?).
The missing data mechanism may vary across patients, and  one can apply the most conservative strategy (i.e. J2R) to patients who drop out due to  lack of efficacy and safety issues. Alternatively, 
one may conduct the tipping point analysis based on the delta-adjusted imputation. 
There are many reasonable ways to assume how the response trajectory changes after treatment discontinuation. 
The MDA algorithm is still suitable as long as the observed data distribution remains the same as that under MAR.

In current clinical practice, it becomes more common to continue the data collection after treatment discontinuation. If  the data observed after treatment discontinuation are assumed to have
 the same distribution as the missing data after dropout, they can be included  in the controlled imputations by using the proposed MDA algorithm with little modifications. 
In the CR and delta-adjusted imputations, we  replace  the assigned  treatment status  by the actual treatment received at each visit (i.e. $0$ after treatment discontinuation) in models  \eqref{factor} and \eqref{factor2}. In J2R and CIR,
 we need to use model \eqref{mmrm_2}, and code the actual  treatment status by $p$ covariates in the covariate set $\mathcal{Z}$.

There are several reasons to implement the controlled imputations  via MI.  The analysis of clinical trials generally follows 
the intention-to-treat  principle \cite{e9:1999, little:1996}, but the data are generated on an as-treated basis \cite{little:1996}.
The MI  inference can accommodate  different imputation and analysis models.  Furthermore,   auxiliary variables and surrogate
outcomes that are correlated with the response variables and the dropout process  may be used to improve  the imputation \cite{tang:2019}. 
Likelihood-based methods have been proposed for the control-based  PMM \cite{lu:2014a}.
As demonstrated in the  supplementary materials of Tang \cite{2017:tangc}, the likelihood-based approach is asymptotically 
equivalent to a MI approach in which both imputation and analysis models follow the as-treated principle, and hence may not be appropriate for the analysis of clinical trials.
Furthermore,  the standard maximum likelihood theory may not work in the SN and ST regressions since the asymptotic distribution for the MLE of the skewness parameters
 can be multimodal for data close to normal \cite{pewsey:2000,Arellano:2008}, and
the MLE of the skewness parameters may be infinite for skewed data \cite{pewsey:2000}.

A variety of multivariate distributions have been proposed for skewed and heavy-tailed data. These include several versions of SN / ST distributions  summarized by Lee and   McLachlan \cite{lee:2013a}, skew-slash distribution \cite{lachos:2009}, skew-contaminated normal distribution \cite{lachos:2009},
and  finite mixtures of  these distributions  \cite{schnatter:2010, lee:2013a}. One popular semiparametric approach  is based on 
the Dirichlet process mixture model \cite{kleinman:1998}, which can be represented as an infinite mixture model.
We choose  the SN and ST distributions developed by   Azzalini  {\it et al} \cite{azzalini:1996,azzalini:2003} because they are easy to work with computationally and effective in handling non-normality for practical purposes.
 The MDA algorithm can be extended to MMRM with residual errors modeled by other non-normal distributions mentioned above. It is also possible to adapt the MDA algorithm as the monotone expectation-maximization algorithm for 
maximum likelihood inferences in these models.

We employ the MMRM model \eqref{mixed} or \eqref{mmrm_2} for missing data imputation, which can be  reorganized as a  sequence of  conditional models \eqref{factor} or \eqref{factor2}. 
We can also build the imputation process directly  on models \eqref{factor} or \eqref{factor2}.
As discussed in Tang \cite{tang:2019},
there are some advantages of using the sequential regression models. First,  there is no need to include all the historical outcomes $(y_{i1},\ldots,y_{ij-1})$ as the predictors of $y_{ij}$ particularly when the number of
response variables $p$ is large. Second, one can incorporate interactions between predictors into the conditional models.

In MMRM-st, the latent variables $(\wc_{i},d_{i})$ are shared by all observations within a subject. 
It  is more parsimonious than the sequential approach based on the univariate ST regression developed in our previous work\cite{tang:2019}, in which $p$ pairs of latent variables $( \wc_{ij},d_{ij})$'s are introduced  per subject.
The sequential ST regression allows the skewness of $y_{ij}$'s to be induced by different latent variables, and the df  to vary by visit / variable, and hence may be more suitable for multivariate data consisting of different  outcomes (e.g. cholesterol, weight) than MMRM-st. 
Although the  sequential ST regression seems more flexible, a large sample size is needed to accurately estimate  \cite{VASCONCELLOS:2005} the df parameters  and detect the difference in the df across visits since the likelihood
function becomes flatter with increasing df.
It may be preferable to use MMRM-st to analyze  longitudinal data with the same response variable collected repeatedly over time
if there is no big variation  in the degrees of tail heaviness  across visits.

Extensive research indicates that the analysis of non-normal outcomes based on the normality assumption may produce  inefficient inferences 
possible because  the violation of normality tends to have more impact on the estimation of the variance-covariance parameters and the variance of the fixed effects 
than on the estimation of the fixed effects \cite{pinheiro:2001}
in both Bayesian inference \cite{lachos:2009} and maximum likelihood inference \cite{pinheiro:2001, zhang:2001,lange:1989}.
This is also observed in the Bayesian analysis of  the cholesterol data using MMRM-n. The Bayes parameter estimates in MMRM-st and MMRM-t are quite insensitive to the outliers.

As evidenced by the DIC criterion, MMRM-st provides the best fit to both the  antidepressant trial and cholesterol data.
In the MI inference, we recommend imputing missing values using MMRM-st, and there is no need to perform a model selection in large confirmatory trials.
A model selection procedure may pick up a wrong model,  inflating the type I error rate  \cite{lukacs:2009, mundry:2009}.
In our early work \cite{tang:2019}, simulation is conducted for the analysis of bivariate continuous and binary outcomes. 
It shows that the MI estimates from the ST regression  have smaller bias and variance than that based on the normal regression for non-normal continuous outcomes,
while the two approaches have almost the same efficiency for normal outcomes. 
 There are numerical evidences that  MMRM-st tends to outperform MMRM-n for non-normal longitudinal continuous outcomes.
Inference based on a reduced model can be misleading if the corresponding assumption does not hold \cite{tang:2019, wang:2015, pinheiro:2001}.
We will  conduct a formal  simulation study to compare these MMRMs after we find enough computational resources, and report the results elsewhere.

A future research direction associated with the  robust MMRMs  is to identify outliers or atypical observations \cite{wang:2015, pinheiro:2001}. This may help us better evaluate the treatment effects (e.g. is 
the effect of the test treatment driven by few subjects?). 
It is inappropriate to remove these atypical observations  from the analysis as  it affects the accuracy and precision of parameter estimates  \cite{lange:1989}.
In MI, we analyze the imputed data by ANCOVA, which may not be robust to a severe deviation from normality \cite{glass:1972}.
The MI inference may be improved by analyzing the imputed data using a robust approach such as the M-estimation \cite{mehrotra:2012}. 
 
\flushleft{ACKNOWLEDGEMENT}\\
We would like to thank the associate editor and two referees for their helpful suggestions that improve the quality of the work.

\appendix
\section{Appendix}
\subsection{Posterior distributions in the MDA algorithm}\label{postst}

\subsubsection{Posterior distribution of $(d_i, \wc_i,$ \texorpdfstring{$\yv$}{y}$_{im} )$ }\label{postwdint}
Let $\tilde{\yv}_{im}=(\wc_i, \yv_{im})$ and $X_{i}=(x_{i1},\ldots, x_{iq}, \yv_{io}')'$.
Let $U_{ij_o}$ and $U_{ij_m}$ be a partition of the $(\Qss+s_i)\times 1$ vector $(\thetav_j,-1,\zerov_{s_i-j}')'$ according to the elements in 
$X_i$ and $\tilde{\yv}_{im}$.  Let  $y_{ij}^{*}=-U_{ij_o}'X_i$ for model \eqref{mixed}.
For model \eqref{mmrm_2}, $y_{ij}^{*}=-U_{ij_o}'X_i-\sum_{k=1}^\Rs \eta_{k} \underline{z}_{ikj}$. The posterior distribution of $(d_i, \wc_i, \yv_{im} )$ is given by
\begin{eqnarray}\label{postwde}
\begin{aligned}
&\pi( d_i,\wc_i,\yv_{im}|\yv_{io},\phiv,\nu)  \propto f(d_i,\wc_i)  \left[\prod_{j=1}^{s_i}\sqrt{d_i\gamma_j}\right] \exp\left[-\sum_{j=1}^{s_i} \frac{d_i\gamma_j (y_{ij}^{*} - U_{ij_m}^{'}\tilde{\yv}_{im}  )^2}{2}\right]\\
&\propto  d_i^{\frac{\nu+s_i+1}{2}-1} 
               \exp\left[ -d_i\frac{b_d+(\tilde{\yv}_{im}-\mu_{w_i})' A_{w_i} (\tilde{\yv}_{im} -\mu_{w_i})}{2} \right]\\
& \propto \left[1+ \frac{\frac{(\wc_i-\mu_{wi1})^2 }{U_{w11}^2b_d/b_a}}{b_a}\right]^{-\frac{b_a+1}{2}} \mathcal{G}\left(d_i| \frac{b_a+1}{2},  \frac{ b_d + (\wc_i-\mu_{wi1})^2 L_{w11}^2}{2}\right)\\
& \qquad \qquad \qquad  N\left(\yv_{im}| \mu_{im}, \frac{U_{w22}U_{w22}'}{d_i}\right)
\end{aligned}
\end{eqnarray}
subject to $\wc_i>0$, where 
  $\tilde{m}= m_{i}+1$, $A_0$ is a $\tilde{m}_i\times \tilde{m}_i$ matrix with $1$ at its $(1,1)$ entry and $0$ elsehwere,
$A_{w_i}=A_0+ \sum_{j=1}^{s_i} \gamma_j U_{ij_m}' U_{ij_m}$, 
the lower triangle matrix $L_{w_i}=\begin{bmatrix} L_{w11} & \zerov\\
                                       L_{w21} & L_{w22}\\\end{bmatrix}$ satisfies $A_{w_i} = L_{w_i}'L_{w_i}$, $L_{w11}$ is a scalar,
 $U_{w_i}=\begin{bmatrix} U_{w11} & \zerov\\
                                       U_{w21} & U_{w22}\\\end{bmatrix}=L_{w_i}^{-1}$,
 $B_{w_i}=\sum_{j=1}^{s_i} \gamma_j U_{ij_m}'y_{ij}^{*}$, 
 $C_{w_i}=U_{w_i}'B_{w_i}$, and $\mu_{w_i}=A_{w_i}^{-1} B_{w_i} =U_{w_i} C_{w_i}=\begin{bmatrix} \mu_{wi1}\\ \mu_{wi2}\\\end{bmatrix} $,
$\mu_{im}= \mu_{wi2} + U_{w21}L_{w11} (\wc_i-\mu_{wi1})$,
$b_a= \nu+o_i$, $ B_{w_i}' A_{w_i}^{-1} B_{w_i} = C_{w_i}'C_{w_i}$ and
$b_d=\nu+\sum_{j=1}^{s_i} \gamma_j y_{ij}^{*2} - C_{w_i}'C_{w_i}$.
In SAS IML, $L_{w_i}$ can  be computed can use the following syntax
\begin{verbatim}
        index=mtilde:1; 
        Lwi = (root(Awi[index,index]))[index,index];
\end{verbatim}

Equation \eqref{postwde} implies that 
we can draw $(\wc_i,d_i,\yv_{im})$ sequentially from 
\begin{eqnarray}\label{postdwy}
\begin{aligned}
d_i^*& \sim \mathcal{G}\left(\frac{b_a}{2},\frac{b_d}{2}\right),\quad \wc_i|d_i^* \sim N^+\left(\mu_{wi1}, \frac{U_{w11}^2}{d_i}\right) ,\\
& \qquad\qquad d_i|\wc_i,d_i^* \sim \mathcal{G}\left(\frac{b_a+1}{2},  \frac{ b_d + (\wc_i-\mu_{wi1})^2 L_{w11}^2}{2}\right), \\
& \quad\,\yv_{im} | d_i,\wc_i,d_i^* \sim N\left(\mu_{im},\frac{U_{w22}U_{w22}'}{d_i}\right). 
\end{aligned}
\end{eqnarray}
The marginal distribution of $\wc_i$ is $t^+\left(\mu_{wi1}, U_{w11}^2\frac{b_d}{b_a},b_a\right)$, but the marginal distribution of $d_i$ is not gamma due to the restriction $\wc_i>0$.

The posterior distribution of $\yv_{im}$ can be equivalently written as 
\begin{equation}\label{postym}
 \yv_{im}| d_i, \wc_i,\yv_{io},\phiv,\nu \sim N\left(\mu_{im}, \frac{V_{y_{\text{m}}}}{d_i}\right),
\end{equation} 
where   $h$ is the index of the first missing observation for subject $i$ in pattern $s$,
 $\tilde{U}_{ij_m}$ is a sub-vector of  $(\thetav_j,-1,\zerov_{s_i-j}')'$ corresponding to the elements in  $\yv_{im}$, $y_{ij}^{**}=y_{ij}^{*}  - \psi_j\wc_i$, 
$V_{y_{\text{m}}}= (\sum_{j=h}^{s} \gamma_j\tilde{U}_{ij_m}\tilde{U}_{ij_m}')^{-1}$  and
$\mu_{im}=\hat{V}_{y_{\text{m}}} \sum_{j=h}^{s} \gamma_j \tilde{U}_{ij_m}y_{ij}^{**}$.

\subsubsection{Posterior distribution of $g$ in step PX1}\label{postg}
The posterior distribution of $g$ with a Harr prior $g^{-1}$ and Jacobian  $g^{n-p}$ is
\begin{eqnarray*}
\begin{aligned}
\text{pos}(g) &\propto g^{n-p} g^{-1}\, \pi\left(Y_m, gd_1,\ldots,g d_n,W, \thetav_1,\frac{\gamma_1}{g},d_{1_\ps},\ldots,\thetav_p,\frac{\gamma_p}{g},d_{p_\ps}, \nu |Y_o\right) \\
& \propto 
  g^{n-p} g^{-1}\prod_{i=1}^n \left\{\left(gd_i\right)^{\frac{\nu+1}{2}-1} \exp\left[ -g\frac{d_i(\nu+\wc_i^2)}{2}\right ] \right\}  \\
& \,\,\, \,\,\,\,\,\,\,\, \prod_{j=1}^p\left\{ \left(\frac{\gamma_j}{g}\right)^{\frac{n_w+2j+r-p-3}{2}}\exp\left[-\frac{\gamma_j\tilde\thetav_j'E_j\tilde\thetav_j }{2g}\right]\right\} \\
&\propto  g^{\frac{n(1+\nu)-p(n_w+r)}{2}-1} \exp\left[-g\frac{\sum_{i=1}^n d_i(\nu+\wc_i^2)}{2}\right] \exp\left[-\frac{\sum_{j=1}^p \gamma_j\tilde\thetav_j'E_j\tilde\thetav_j }{2g}\right].
\end{aligned}
\end{eqnarray*}

\subsubsection{Posterior distribution of $h$ in step PX2}\label{posth}
The posterior distribution of $h$  with a Harr prior $h^{-1}$ and Jacobian  $h^{n-p}$  is 
\begin{eqnarray*}
\begin{aligned}
\text{pos}(h)\propto  h^{n-p} h^{-1}  \exp\left(-h^2 \frac{\sum_{i=1}^n d_i \wc_i^2}{2}\right) \exp\left(-\frac{\sum_{j=1}^p\gamma_j\underline\psi_j^2 \frac{ 4d_{j_\ps}  }{\pi^2}}{2h^2}\right).
\end{aligned}
\end{eqnarray*}
The posterior distribution of $H=h^2$ is 
$$\text{pos}(H) \propto \text{pos}(h) \left|\frac{\partial h}{\partial H}\right|\propto H^{\frac{n-p}{2}-1} \exp\left(-H \frac{\sum_{i=1}^n d_i \wc_i^2}{2}\right) \exp\left(-\frac{\sum_{j=1}^p\gamma_j\underline\psi_j^2 \frac{ 4d_{j_\ps}  }{\pi^2}}{2H}\right).$$

\subsubsection{Prior and posterior distributions of $\nu$}\label{prepostnu}
We firstly derive the PC prior for $\nu$. The Kullback-Leibler (KL) distance between the multivariate t distribution $t(\muv, \frac{\nu-2}{\nu}\Sigma, \nu)$ 
and the normal distribution $N(\muv,\Sigma)$ is
\begin{eqnarray*}
\begin{aligned}
 \text{KL}(\nu) &=\int t\left(\xv|\muv,\frac{\nu-2}{\nu}\Sigma,\nu\right)\log t\left(\xv|\muv,\frac{\nu-2}{\nu}\Sigma,\nu\right)d\xv  \\
& \,\,\,\,\,\,\,\,   \,\,\,\,\,\,\,\,       - \int t\left(\xv|\muv,\frac{\nu-2}{\nu}\Sigma,\nu\right) \log \phi(\xv|\muv,\Sigma)d\xv \\
&=
 \frac{p}{2}\left[1+\log\left(\frac{2}{\nu-2}\right)\right] +\log\Gamma\left (\frac{\nu+p}{2}\right)-\log\Gamma \left(\frac{\nu}{2}\right) \\
&  \,\,\,\,\,\,\,\,   \,\,\,\,\,\,\,\,          -\frac{\nu+p}{2}\left[\Psi\left(\frac{\nu+p}{2}\right)-\Psi\left(\frac{\nu}{2}\right)\right].
\end{aligned}
\end{eqnarray*}
since the first integration equals $\log\Gamma \left(\frac{\nu+p}{2}\right)-\log\Gamma \left(\frac{\nu}{2}\right) -\frac{\nu+p}{2}\left[\Psi\left(\frac{\nu+p}{2}\right)-\Psi\left(\frac{\nu}{2}\right)\right]-\frac{1}{2}\log|\Sigma|
  -\frac{p}{2}\log(\nu-2)-\frac{p}{2}\log(\pi)$ by Kotz and Nadarajah \cite{kotz:2004}, and the second integration equals  $-\frac{p}{2}\log(2\pi)-\frac{1}{2}\log|\Sigma|-\frac{p}{2}$.

 By the definition of the PC prior \cite{simpson:2017}, the density is $\lambda \exp(-\lambda d(\nu))\left| \frac{\partial d(\nu)}{\partial{\nu}}\right|$, where $d(\nu)=\sqrt{2 \text{KL}(\nu)}$.

 The posterior distribution of $\nu$ is given by
\begin{eqnarray*}
\begin{aligned}
&  \pi(\nu|\Sigma, \psiv,\theta_j,\gamma_j) \propto \pi(\nu) \prod_{i=1}^n  t_p (\yv_{io}|\muv_{io},\Omega_{io},\nu) \\
& \,\,\,\,\,\,\,\,\,\, \,\,\,\,\,\,\,\,\,\,  T_{\nu+o_i} \left[\lambdav_{io}^{*'} (\yv_{io}-\muv_{io})\sqrt{\frac{\nu+o_i}{\nu+(\yv_{io}-\muv_{io})'\Omega_{io}^{-1}(\yv_{io}-\muv_{io})}}\,\right]I(\nu>\nu_l),
\end{aligned}
\end{eqnarray*}
where $t_p (\yv_{io}|\muv_{io},\Omega_{io},\nu) \propto \frac{\Gamma(\frac{\nu+o_i}{2})}{\Gamma(\frac{\nu}{2}) \nu^{o_i/2}} [1+(\yv_{io}-\muv_{io})'\Omega_{io}^{-1}(\yv_{io}-\muv_{io})/\nu]^{-\frac{\nu+o_i}{2}}$, and $\lambdav_{io}^*= \Sigma_{io}^{-1}\psiv_{io} / \sqrt{1+\psiv_{io}' \Sigma_{io}^{-1}\psiv_{io}}$. 

For subjects with no intermittent missing data, the skew-t density function  can be computed without matrix inversion using the following relationship
 $\lambdav_{io}^{*'} (\yv_{io}-\muv_{io})=\sum_{t=1}^s \lambda_t R_t\underline\psi_t$ and 
$(\yv_{io}-\muv_{io})'\Omega_{io}^{-1}(\yv_{io}-\mu_{io})= \sum_{t=1}^s \lambda_t r_t^2 -[\sum_{t=1}^s\lambda_t r_t\underline\psi_t]^2/ [1+ \sum_{t=1}^s \lambda_t \underline\psi_t^2]$, where
$r_j=y_{ij}- \sum_{t=1}^{j-1}\beta_{jt}y_{it}  - \sum_{k=1}^{\Qs} \underline\alpha_{kj} x_{ik}$ for model \eqref{mixed} and
$r_j=y_{ij}- \sum_{t=1}^{j-1}\beta_{jt}y_{it} -\sum_{k=1}^{\Rs} \eta_{k} \underline{z}_{ikj} - \sum_{k=1}^{\Qs} \underline\alpha_{kj} x_{ik}$ for model \eqref{mmrm_2}.  

The candidate $\nu^*$ is generated from $\log(\nu^*-\nu_l)\sim N[\log(\nu-\nu_l), c^2]$.
It will be accepted with probability $\alpha_\nu=\min\left\{1, \frac{(\nu^*-\nu_l) \pi(\nu^*)\prod_{i=1}^n f(\yv_{io}|\thetav_i,\gamma_i,\nu^*) }{(\nu-\nu_l) \pi(\nu)\prod_{i=1}^n f(\yv_{io}|\thetav_i,\gamma_i,\nu)}\right\}$.

\subsection{Adaption of the MDA algorithm for MMRM-sn}\label{adaptsn}

For MMRM-sn (i.e. $\nu\equiv \infty$), the MDA algorithm $A$ can easily adapted by ignoring steps P2 and PX1, and setting $d_i\equiv 1$ and $d_i^*\equiv 1$ in drawing 
$(\thetav_j,\gamma_j,d_{\psi_j}, \rho_j)$'s and $(\wc_i,\yv_{im})$'s. For example, 
$(\wc_i, \yv_{im})$  in step I can be imputed by modifying the posterior distribution \eqref{postdwy} as
$$\wc_i \sim N^+(\mu_{wi1}, U_{w11}^2) \,\text{ and } \,\yv_{im} | \wc_i \sim N(\mu_{im}, U_{w22}U_{w22}').$$

\subsection{Adaption of the MDA algorithm for MMRM-t}\label{adaptt}
For MMRM-t (i.e. $\underline\psi_j\equiv 0$ for $j=1,\ldots,p$, $\wc_i\equiv 0$), the MDA algorithm $A$ needs the following modifications:\\
1. Step PX2 is no longer needed. \\
2. Step P1: Remove $\wc_i$ from the model. Set $d_{\psi_j}\equiv 0$ and $\thetav_j=(\underline\alpha_{1j},\ldots,\underline\alpha_{q j},\beta_{j1},\ldots,\beta_{jj-1})'$.
Sample $(\thetav_j,\gamma_j)$'s  from
\begin{eqnarray*}
\begin{aligned}
 \pi(\thetav_j, \gamma_j|d_i\text{'s},\nu,Y_o, Y_m) \propto \gamma_j^{\frac{n_w+2j+r^*-p-1}{2}-1} \exp\left[-\frac{\gamma_j}{2} \tilde\thetav_j' (E_j+\sum_{i\leq n_j} d_i\tilde{\xv}_{ij}\tilde{\xv}_{ij}') \tilde\thetav_j\right] 
\end{aligned}
\end{eqnarray*}
for $j=1,\ldots,p$,
where $E_j$ is the $(q+j)\times (q+j)$ leading principle submatrix of the $(q+p)\times (q+p)$  matrix $E=  \begin{bmatrix}  M  & M\alphav_0'  \\
                                 \alphav_0 M &  \alphav_0 M \alphav_0' +A_w\\ \end{bmatrix} $ and $r^*$ is the rank of $M$. If the inverse Wishart or Jeffrey's prior (with fixed $A_w$ and $n_w$) instead of the hierarchical prior of Huang and Wand \cite{huang:2013} is used, step P0 shall be ignored.\\
3. Step I: draw $d_i\sim  \mathcal{G}\left(\frac{b_a}{2},\frac{b_d}{2}\right)$ and $\yv_{im} \sim N(\mu_{w_i}, (d_iA_{w_i})^{-1})$ since
\begin{eqnarray*}
\begin{aligned}
 \pi( d_i, & \yv_{im}|\yv_{io}, \nu,\phiv)  \propto  f(d_i)  d_i^{\frac{s_i}{2}}\exp\left[- \frac{\sum_{j=1}^{s_i}d_i\gamma_j (y_{ij}^{*} - \tilde{U}_{ij_m}'\yv_{im})^2}{2}\right] \\
\propto &  \left\{ d_i^{\frac{\nu+o_i}{2}-1} \exp\left[ -\frac{d_ib_d}{2}\right]\right\}
              \left\{ d_i^{\frac{m_i}{2}}  \exp[ -\frac{d_i(\yv_{im} -\mu_{w_i})' A_{w_i} (\yv_{im} -\mu_{w_i})}{2} ]\right\},
\end{aligned}
\end{eqnarray*}
 where $f(d_i)\propto d_i^{\nu/2-1}\exp(-d_i\nu)$,
$A_{w_i}= \sum_{j=1}^{s_i} \gamma_j \tilde{U}_{ij_m}' \tilde{U}_{ij_m}$,  $B_{w_i}=\sum_{j=1}^{s_i} \gamma_j \tilde{U}_{ij_m}'y_{ij}^{*}$,  $\mu_{w_i}=A_{w_i}^{-1} B_{w_i}$,
$b_a= \nu+o_i$ and
$b_d=\nu+\sum_{j=1}^{s_i} \gamma_j y_{ij}^{*2} -  B_{w_i}' A_{w_i}^{-1} B_{w_i}$.\\
4. Step PX1: $g$ is randomly drawn from \\
$g^{\frac{n\nu-p(n_w+r^*)}{2}-1} \exp\left[-g\frac{\nu\sum_{i=1}^n d_i}{2}\right] \exp\left[-\frac{\sum_{j=1}^p \gamma_j\tilde\thetav_j'E_j\tilde\thetav_j }{2g}\right]$.

 \bibliographystyle{SageV} 
\bibliography{mda_skewt}

\end{document}